\documentclass[pdftex,onecolumn,epjc3]{svjour3}          
\RequirePackage[T1]{fontenc}
\pdfoutput=1
\usepackage{tikz}
\usepackage{tikz-cd}

\usetikzlibrary{arrows,decorations.markings}

\usepackage{amssymb}
\usepackage{amsmath}
\usepackage{amsbsy}    
\RequirePackage{graphicx}
\RequirePackage{mathptmx}      
\RequirePackage{flushend}
\RequirePackage[numbers,sort&compress]{natbib}
\RequirePackage[colorlinks,citecolor=blue,urlcolor=blue,linkcolor=blue]{hyperref}
\def\bea{\begin{eqnarray}}
\def\eea{\end{eqnarray}}
\usepackage[titletoc,title]{appendix}
\usepackage{blindtext}

\usepackage{mathtools}
\usepackage{mathrsfs}
\usepackage{amsfonts}
\usepackage{enumerate}
\usepackage{array}
\usepackage{booktabs}
\usepackage{textcomp}
\usepackage{caption}
\usepackage{cancel}
\usepackage{units}
\usepackage{multirow}
\usepackage{pst-all}
\usepackage{longtable}
\usepackage{relsize}
\usepackage{natbib}
\usepackage{amsfonts}
\usepackage[T1]{fontenc}
\usepackage[utf8]{inputenc}
\usepackage{hyperref}

\newcommand{\squeezeup}{\vspace{5.5mm}}
\journalname{Eur. Phys. J. C}
\usepackage{etoolbox}
\usepackage{enumitem}

\usepackage[margin=0.8in]{geometry}

\begin{document}
\setlength{\abovedisplayskip}{0pt}
\setlength{\belowdisplayskip}{0pt}

\title{Topological Strings on Toric geometries in the presence of Lagrangian branes }

\author{M.Nouman Muteeb
\thanksref{e1}
}

\thankstext{e1}{e-mail: nouman01uet@gmail.com}

\institute{Abdus Salam School of Mathematical Sciences, Lahore, Pakistan}
\date{Received: date 
/ Accepted: date
}
\maketitle
\begin{abstract}
We propose expressions for refined open topological string partition function on certain non-compact Calabi Yau 3-folds with topological branes  wrapped on the special lagrangian submanifolds. The corresponding  web diagrams are partially compact and a lagrangian brane is inserted on one of the external legs. Partial compactification introduces a mass deformation in the corresponding gauge theory. We propose conjectures that equate these open topological string partition functions with the generating function of equivaraint indices on certain quiver moduli spaces. To obtain these conjectures we use the identification of topological string partition functions with equivariant indices on the instanton moduli spaces. 
\end{abstract}
\tableofcontents
\newpage
\section{Introduction}
Topological strings  \cite{Antoniadis:1993ze,Neitzke:2004ni,Assi:2014exa} correspond to the subsector of  full superstrings spectrum that is invariant under a non-trivial sub algebra of the extended supersymmetry algebra. The  observables of this subsector  describe various topological properties of the spacetime on which the observables are defined. In physical  4d  effective field theory  the topological strings with the so-called $A-twist$ generate the the following F-term that depends on the vector multiplet moduli
\bea
\int F_g(\mathcal{W}^2)^g
\eea
where $\mathcal{W}^2$ is composed of the Weyl multiplet superfield $\mathcal{W}_{\alpha\beta}$ and $F_g$ is the genus $g$ topological string free energy. $F_g$ for $g\ge 2$ describe coefficients in the scattering amplitude of $2g-2$ graviphotons.
The presence of D-branes imply  consistent boundary conditions on the world sheet boundaries. In the topological sector the boundary conditions should preserve the BRST symmetry of the world sheet. From the  target space (CY)  perspective in the A-model the world sheet boundaries are mapped to a particular submanifold $L$ whose dimension is  equal to the half that of CY \cite{Neitzke:2004ni} and the restriction of the CY K\"ahler form $\omega$ to $L$ vanishes. In general, open strings may end on $L$ resulting in the wrapping of an A-model brane on L. 
\\
The problem of computing the unrefined topological string amplitudes on the toric CY in the presence of both  external as well as internal branes  was solved by the technique of the topological vertex \cite{Aganagic:2003db} .  On the other hand the refined topological string amplitude in the presence of   internal branes turns out to be a subtle problem.
Certain surface operators in 4d gauge theory can be realised by wrapping D4-branes on the Lagrangian 3-cycles of the CY 3-fold with the other two directions extending along transverse $\mathbb{R}^2\subset\mathbb{R}^4$. These two transverse directions correspond to equivariant parameters  $\epsilon_1,\epsilon_2$ or $t=e^{-i\epsilon_1},q=e^{i\epsilon_2}$. 
The topological branes can be put on the external non-compact legs or the internal compact legs of the toric diagram associated to the CY. 
Progress has been made in \cite{Kozcaz:2018ndf}, where   authors discuss  refinement of holonomies for refined open topological string amplitudes. This was the case for topological branes on the external leg of the toric  web diagram.
In the presence of  internal brane(s), there is a mismatch between the results as computed by using refinement of holonomy prescription and the geometric transition.\\
In this note we suggest  that by utilising the equivalence of topological string amplitudes with equivariant indices \cite{Li:2004ef} on framed moduli spaces, it may be possible to compute the refined open string amplitudes in the presence of internal branes. To this end, the authors
  \cite{Bruzzo:2010fk}  proposed a conjecture that equates the refined open  string invariants of the special lagrangian branes in toric Calabi Yau 3-folds, with the Witten index of  the supersymmetric quantum mechanical model describing the BPS states attached to the surface defect.  The Witten index can  be interpreted as the Euler characteristic of the moduli space described by the quantum mechanical model. In other words the conjecture \cite{Bruzzo:2010fk}  equates the generating function of  refined open  string invariants of the special lagrangian branes with the  instanton partition function in the presence of surface defects. The conjecture was checked  to be true to high orders in the asymptotic expansion. According to the geometric engineering argument the $M5$ branes wrapped on a  submanifold of the CY give rise to effective five dimensional gauge theories with surface defect.  \\
In the case of partially compactified toric web diagram the  refined open topological string amplitude  is equated to the generating function of $\chi_y$ genus on the same moduli space \cite{Li:2004ef,Hollowood:2003cv,Chuang:2013wpa}.  The K\"ahler parameter of this new compact direction corresponds to a massive adjoint hyper in the 5d gauge theory \cite{Hollowood:2003gr}. We  generalise the results of  {\cite{Bruzzo:2010fk} and state the conjectures in the case of partial compactification of the resolved conifold, $O(-1) \oplus O(-1) \to \mathbb{P}^1$ and the partial compactification of  the total space of the canonical bundle $\mathcal{O}(-2, -2)$ of $P^1 \times P^1$. In the  fully compactified case we give the expression for the generating function of   elliptic genus of the defect moduli space and speculate about its relevance for  the refined open topological string amplitude in the presence of internal branes. \\
In section (\ref{section2}) we restate the conjecture proved in \cite{Bruzzo:2010fk}, about the equivalence of partition function on  quiver moduli space with open Gromov-Witten invariants of certain non-compact Calabi-Yau 3-folds. In section (\ref{section:chiy5d}) we compute the generating function of $\chi_y$ genus on the quiver moduli space. The expression we obtain is not a polynomial in the mass-deformation parameter $y$.
In the next section (\ref{anylcont}) we use analytic continuation to write the $\chi_y$-genus as a polynomial in $y$. In section \ref{sectionr2} we compute the generating function $\chi_y$-genus for quiver moduli space for  rank $r=2$, which corresponds to the refined open topological string partition function on the partially compactified web of the Hirzebruch surface $F_0$. In section (\ref{DTKT}) we briefly discuss the identification of the Donaldson-Thomas partition function of CY3-fold and K-theory partition function of 5d  supersymmetric gauge theory motivated by the geometric engineering argument. This identification was obtained for the unrefined case by making a change of variables. Using certain consistency conditions we propose a generalisation of this change of variables to the refined case.  In the last section (\ref{elliptge2}) we given an expression for the elliptic genus on the same moduli space and suggest that it may be related to the open topological string partition function on non-compact  Calabi-Yau 3-fold whose corresponding   web diagram is fully compactified.The appendices (\ref{appendix:A},\ref{appendix:B},\ref{appendix:C})  contain the refined topological vertex amplitudes for the remaining preferred directions and the appendix (\ref{virtuallocalization}) contain a summary of the virtual equivariant localisation and fixed point theorems.
\section{Quiver model and Open topological string amplitudes}\label{section2}
A standard   D-brane construction of  $5d$  gauge theory with eight supercharges  involves \cite{Bruzzo:2010fk} $D6$-branes wrapped on the holomorphic curves
 in a non-compact $K3$ surface $S$. 
  The type of quiver is defined by the intersection matrix of the configuration of $(-2)$-rational curves after suitable resolution. In the present context only $A_r$-type singularities are considered which are amenable to analysis with toric geometry techniques.  So the type IIA vacuum we consider is given by $S\times S^1\times\mathbb{R}^5\equiv S\times C^*\times\mathbb{R}^4$. The low energy limit of $D6$-branes wrapped on these rational curves gives rise to $5d$  quiver gauge theory with eight supercharges. In the more general type IIA superstrings setup we can add D4-branes wrapping the special Lagrangian submanifolds, along with D2 branes ending on these D4-branes. D2-branes are wrapped on  $(-2)$-rational curves. In other words D4-branes serve as defect operators and D2 branes correspond to the BPS states bound to the defect operators. The world volume of  D4-brane is $L\times \mathbb{R}^2$, where $L\subset T\times C^*$ is the special Lagrangian submanifold and $\mathbb{R}^2\subset \mathbb{R}^4$. By construction a toric Calabi Yau 3-fold $X$ admits a symplectic $U(1)^3$ action and the resulting moment map $\rho:X\to \mathbb{R}^3$ to the so-called Delzant polytope. The collection of $U(1)^3$ preserving compact and non-compact rational holomorphic curves of $X$ define its  toric skeleton. The toric skeleton is mapped by $\rho$ to a trivalent graph $\Delta$ in $\mathbb{R}^3$. The special lagrangian submanifold $L$ under consideration is topologically equivalent to $S^1\times \mathbb{R}^2$ and  is  mapped to a half real line which intersects a 1-face of the graph $\Delta$. The external lagrangian cycles intersect the non-compact components of the skeleton, whereas the internal lagrangian cycles intersect compact components.  In the web diagrams we always show the branes wrapped on the lagrangian cycles by dashed lines.
\\
The D4-branes are wrapped on the Lagrangian cycles of the CY3-fold, with two directions extending along one of the $\mathbb{R}^2s$ of the  transverse $ \mathbb{R}^4$. For the unrefined case the choice of $\mathbb{R}^2$ is immaterial. For the refined case the two $\mathbb{R}^2$s inside $\mathbb{R}^4$ are rotated by $q=e^{i\epsilon_1},t=e^{-i\epsilon_2}$ corresponding to a particular choice of complex structure. Depending on which $\mathbb{R}^2$ the D4-brane is extended along, it is called either a q-brane or a t-brane \cite{Kozcaz:2018ndf}.
In the presence of D4-branes  the open topological string amplitudes are the generating functions of BPS degeneracies of D2 branes. These D2-branes are wrapped on smooth curves whose boundaries lie on D4-branes. The boundary conditions are necessary for  the complete specification of the open string amplitudes and are given by gauge invariant combinations of  holonomy operators.
\\
The pure $SU(r),r\ge 2$ gauge theories in 5d can be engineered by certain non-compact CY 3-folds. To construct  these 3-folds , the total space of $\mathcal{O}(-1)\oplus \mathcal{O}(-1)\to \mathbb{P}^1$  is orbifolded by the action  $(z_1,z_2)\to (e^{\frac{2\pi i}{r}}z_1,e^{-\frac{2\pi i}{r}}z_2)$ on the fiber coordinates $(z_1,z_2)$. The resulting space is singular and can be resolved to a smooth CY 3-fold which contains $(r-1)$  geometrically ruled surfaces glued together. It is equivalent to the resolved $A_{r-1}$ fibration over $\mathbb{P}^1$. The compact part of the geometry consist of $r-1$ Hirzebruch surfaces glued together  and the normal geometry of a base $\mathbb{P}^1$ in the $p-1$-th and $p$-th Hirzebruch surfaces is $\mathcal{O}(-r+2p-2)\oplus \mathcal{O}(r-2p)$. Formally allowing the value $r=1$ corresponds to the resolved conifold, the total space of $\mathcal{O}(-1)\oplus \mathcal{O}(-1)\to \mathbb{P}^1$. \\  
Partially compactifying the web diagrams, which becomes non-planar, changes the CY3-fold to an elliptic CY3-fold which has the structure of the form $\mathbb{C}^2/\mathbb{Z}_r\times_f T^2$. M-theory compactification on this geometry engineers $5d$ $\mathcal{N}=1^*$ gauge theory with a single adjoint hypermultiplet \cite{Hollowood:2003cv,Iqbal:2008ra}.\\
The fields of quantum mechanical system, describing the BPS states bound to the surface operators, arise from low energy modes of  $D2-D6$, $D2-D4$ and $D2-D2$ configuration of D-branes in type IIA strings. The field content is summarised as 
\bea
D2-D6&:& two\quad  (0,2)\quad chiral\quad multiplets\nonumber\\
 D2-D4&:& a\quad (0,2)\quad chiral\quad multiplet\quad and\quad a\quad (0,2)\quad vector\quad\nonumber\\ &.&multiplet\nonumber\\
 D2-D2^{\prime}&:&two\quad (0,2)\quad chiral\quad multiplets\quad and\quad two\quad (0,2)\quad\nonumber\\ &.&Fermi\quad multiplets\nonumber\\
D2^{\prime}-D6&:&  a\quad single \quad (0,2)\quad Fermi \quad multiplet.
\eea
\\
 An intricate  analysis \cite{Bruzzo:2010fk} shows that the moduli space of the supersymmetric vacua of the quantum mechanical model is isomorphic to the data defining  ADHM type quiver. In a certain region of the moduli space it is interpreted in terms of certain generalised vector bundles. \\
  The D2-brane effective action is derived by the dimensional reduction of the field contents of quiver diagram (\ref{D2quiver}). The quiver diagram describes a vector space of supersymmetric flat directions parametrised by the fields $(A_1,A_2,I,J,B_2,f,g,\sigma_1,\sigma_2)$ as follows
\bea
&\mbox{End}&(V_1)^{\oplus2}\oplus \mbox{Hom}(W,V_1)\oplus \mbox{Hom}(V_1,W)\oplus \mbox{End}(V_2)\oplus \mbox{Hom}(V_1,V_2)\oplus \mbox{Hom}(V_2,V_1)\nonumber\\&\oplus& \mbox{u}(V_1)\oplus \mbox{u}(V_2)\nonumber\\
\eea
for hermitian inner product vector spaces $V_1,V_2$ and $W$. The vacuum equations of the D2-branes effective action as given below define a moduli space
\bea\label{eq:flat}
&\hspace{0.1cm}&[A_1,A_1^{\dagger}]+[A_2,A_2^{\dagger}]+I I^{\dagger}-J^{\dagger}J+ff^{\dagger}-g^{\dagger}g=\zeta_1,\qquad
[B_2,B_2^{\dagger}]-f^{\dagger}f-g g^{\dagger}=\zeta_2,\nonumber\\ 
&\hspace{0.1cm}&gA_1=0,\quad A_1 f=0,\quad gI=0,\quad Jf=0\quad [A_1,A_2]+IJ=0
A_2f-fB_2=0,\nonumber\\&\hspace{0.1cm}& gA_2-B_2g=0, fg=0,\nonumber\\
&\hspace{0.1cm}&[ \sigma_1 , A_1 ]=0,\quad [ \sigma_1 , A_2 ]=0,\quad [ \sigma_2 , B_2 ]=0,\quad \sigma_1I=0,\quad J\sigma_1=0,\quad \sigma_1f-f\sigma_2=0,\nonumber\\ &\hspace{0.1cm}&g\sigma_1-\sigma_2g=0.\nonumber\\
\eea
The moduli space parametrize  $\mbox{U}(V_1)\times \mbox{U}(V_2)$ gauge inequivalent solutions to (\ref{eq:flat}). An important result proven in \cite{Bruzzo:2010fk} shows that the moduli space defined by the last set of equations is isomorphic to a different moduli space for generic values of Fayet Illuopolous parameters.The later moduli  space comprises of stable representations of the enhanced ADHM quiver, also described in terms of framed torsion free sheaves on the projective plane.
The virtual smoothness in our context means that  moduli space  of stable representations of the ADHM  quiver can be embedded in a smooth variety which is a hyper k\"ahler quotient. 

\begin{figure}
\begin{tikzpicture}
\begin{tikzcd}
\end{tikzcd}
\end{tikzpicture}
\end{figure}
\begin{figure}
\begin{tikzpicture}
\begin{tikzcd}
\end{tikzcd}
\end{tikzpicture}
\end{figure}
\begin{figure}
\begin{tikzpicture}
\begin{tikzcd}
\end{tikzcd}
\end{tikzpicture}
\end{figure}
\begin{figure}
\begin{tikzpicture}
\begin{tikzcd}
\arrow[out=175,in=240,loop,swap,"B_{2+}"]
V_2 \arrow[r,bend left,"\Phi_+\Omega_-"] 
\arrow[r,bend right,swap,"\Gamma_+\Psi_-"] & V_1
 \arrow[out=50,in=100,loop,swap,"A_{1+}A_{2+}"]
 \arrow[out=250,in=300,loop,swap,"\chi_-"]
  \arrow[r,bend left,"\mathcal{I}"] \arrow[r,bend right,swap,"\mathcal{J}"] & W
\draw[ thick, ->] (0,0) arc (0:183:1.6);
\node [right] at (-0.6,1) {$\Lambda_-$};
\end{tikzcd}
\end{tikzpicture}
\caption{quiver diagram encoding D2-branes effective action}
    \label{D2quiver}
\end{figure}
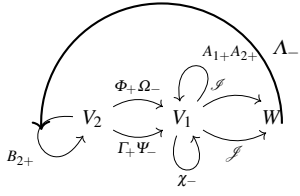
The enhanced ADHM quiver  is given in figure (\ref{enquiver}) along with the $relations$
\bea\label{relations}
&\alpha_!\alpha_2-\alpha_2\alpha_1+\xi\eta,\quad \alpha_1\phi,\quad \alpha_2\phi-\phi\beta,\quad \eta\phi,
\quad \gamma\xi\quad,\phi\gamma,\gamma\alpha_1,\gamma\alpha_2-\beta\gamma.
\eea
A triple of vector spaces $(V_1,V_2,W)$ is assigned to the vertices $(e_1,e_2,e_{\infty} )$ as a representation. Similarly the linear maps $(A_1,A_2,I,J,B,f,g)$ represent the arrows $(\alpha_1,\alpha_2,\xi,\eta,\beta,\phi,\gamma)$. The resulting quiver representation is moded out by the relations (\ref{relations}). If we define $(r,n_1,n_2):=(\mbox{dim}(W),\mbox{dim}(V_1),\mbox{dim}(V_2))$, then this triple of positive integers define the  numerical type of the quiver. The framing of the  enhanced quiver representation corresponds to the existence of an isomorphism $\mbox{h}:\mbox{W}\to \mathbb{C}^r$. Moreover a stable\footnote{more precisely $\theta$-semistable} representation of type $(r,n_1,n_2)$ implies the existence of a set of parameters $\theta=(\theta_1,\theta_2,\theta_{\infty})$ satisfying the relation $n_1\theta_1+n_2\theta_2+r\theta_{\infty}$ such that
\begin{itemize}
\item Any subrepresentaion of type $(0,m_1,m_2)$ satisfies $m_1\theta_1+m_2\theta_2\le0$.
\item Any subrepresentaion of type $(r,m_1,m_2)$ satisfies $m_1\theta_1+m_2\theta_2+r\theta_{\infty}\le0$
\end{itemize}
The following result gives criterion for generic stability conditions:\\
Given a quiver representation $\mathcal{R}$ of type $(r,n_1,n_2)$ and theta parameters satisfying $\theta_2>0,\theta_1+n_2\theta_2<0$, then the following three statements are equivalent \\
\begin{itemize}
\item1. $\mathcal{R}$ is $\theta$-semistable
\item2.$\mathcal{R}$ is $\theta$-stable
\item3. (a) $f:V_2 \to V_1$ is injective and $g:V_1\to V_2$ is identically zero\\
(b) The data $\mathcal{A}=(V_1,W,A_1,A_2,I,J)$ satisfies the ADHM stability conditions. 
\end{itemize}
Consider the vector spaces $V_1,V_2,W$ with positive definite  dimensions $n_1,n_2,r$ and the direct sum of vector spaces
\bea
X(r,n_1,n_2)=\mbox{End}(V_1)^{\oplus 2}\oplus \mbox{Hom}(W,V_1)\oplus \mbox{Hom}(V_1,W)\oplus \mbox{End}(V_2)\oplus \mbox{Hom}(V_1,V_2)\oplus \mbox{Hom}(V_2,V_1)\nonumber\\
\eea
$X(r,n_1,n_2)$ admits a $\mbox{GL}(V_1)\times \mbox{GL}(V_2)$ action defined as
\bea
(g_1,g_2)\times(A_1,A_2,I,J,B_2,f,g)\to (g_1A_1g_1^{-1},g_1A_2g_1^{-1},Jg_1^{-1},g_1I,g_2B_2g_2^{-1},g_1fg_2^{-1},g_2gg_1^{-1})\nonumber\\
\eea
It is clear that $\mathcal{G}=\mbox{GL}(V_1)\times \mbox{GL}(V_2)$ preserves the subset $X_0(r,n_1,n_2)\in X(r,n_1,n_2)$ defined  by (\ref{eq:flat}). The space $X_0(r,n_1,n_2)$ parametrises the representations $\mathcal{R}=(V_1,V_2,W,A_1,I,J,B_2,f,g)$ with two framed representations $\mathcal{R}_1,\mathcal{R}_2$ being equivalent if the corresponding points in $X_0(r,n_1,n_2)$ belong to the same $GL(V_1)\times GL(V_2)$ orbit.
The space $X_0(r,n_1,n_2)$ can be  projectivized to a scheme as follows
\bea
\mathcal{N}_{\theta}^{ss}(r,n_1,n_2)=X_0(r,n_1,n_2)//_{\chi}\mathcal{G}:=Proj\bigg(\oplus_{n\ge 0}A(X_0(r,n_1,n_2))^{G,\chi^n}  \bigg)
\eea 
with the notation $A(X_0(r,n_1,n_2))^{\mathcal{G},\chi^n}:=\{f\in A(X_0(r,n_1,n_2))|f(g.x)=\chi(g)^nf(x)\forall g\in G\}$. There exists an open subscheme $\mathcal{N}_{\theta}^{s}(r,n_1,n_2)$ of $\mathcal{N}_{\theta}^{ss}(r,n_1,n_2)$ whose $\mathcal{G}$-orbits are $\chi$-stable\footnote{The character  function $\chi:\mathcal{G}\to \mathbb{C}^{\times}$, where $\mathcal{G}$ acts on the space $X(r,n_1,n_2)$, furnishes a definition of $\chi$-stability. For this consider the existence of a polynomial $q(x)$ on  $X(r,n_1,n_2)$ such that $q((g_1,g_2).x)=\chi(g_1,g_2)^nq(x)$ for positive definite integer $n$. If $q(x_0)\ne 0$, $x_0$ is called $\chi$-semistable. Moreover if  $\Delta\subset \mathcal{G}$ acts trivially on $X(r,n_1,n_2)$ such that $dim(\mathcal{G}.x_0)=dim(\mathcal{G}/\Delta)$  and the action of $\mathcal{G}$ on all such $x_0$ is closed, then then $x_0$ is called $\chi$-stable.}. In the parameter space defined by the inequalities $\theta_2>0,\theta_1+n_2\theta_2<0$ the framed representations of the enhanced quiver that satisfy condition $(3)$ can simply be denoted by $\mathcal{N}(r,n_1,n_2)$ by dropping subscripts and superscripts.\\
The matter couplings in the quantum mechanical model corresponds to the three tautological bundles $\mathcal{V}_1,\mathcal{V}_2,\mathcal{W}$ on the moduli space $\mathcal{N}(r,n_1,n_2)$. These bunldes are defined by $\mathcal{W}=\mathcal{O}^{\oplus r}_{\mathcal{N}(r,n_1,n_2)}$, $\mathcal{L}_1=\mbox{det}(\mathcal{V}_1)$ and $\mathcal{L}_2=\mbox{det}(\mathcal{V}_2)$. Moreover several copies of $\mathcal{L}_1,\mathcal{L}_2$ can be tensored to give rise to the mixed line bundles $\mathcal{L}_{p_1,p_2}=\mathcal{L}_1^{\otimes p_1}\otimes \mathcal{L}_2^{\otimes p_2}$.\\
 The existence of the  morphism $s:\mathcal{N}(r,n_1,n_2)\to \mathcal{M}(r,n_1-n_2)$, where $ \mathcal{M}(r,n_1-n_2)$ is the moduli space of ADHM data of type $(r,n)$,  plays a simplifying role in the application of the equivariant fixed point theorems. The  tangent space at a point of  $\mathcal{N}(r,n_1,n_2)$ is isomorphic to the difference $H^1(\mathcal{C}(\mathcal{R}))-H^2(\mathcal{C}(\mathcal{R}))$ for $\mathcal{R}=(A_1,A_2,I,J,B_2,f)$ representing a point of $\mathcal{N}(r,n_1,n_2)$, where the complex $\mathcal{C}(\mathcal{R})$ is defined by
 \begin{figure}
 \[
\begin{tikzcd}
\mbox{End}(V_1)\oplus \mbox{End}(V_2)  \arrow[r,"d_0"] & \mbox{End}(V_1)^{\oplus 2} \oplus \mbox{Hom}(W,V_1)\oplus \mbox{Hom}(V_1,W)\oplus \mbox{End}(V_2)\oplus \mbox{Hom}(V_2,V_1) 
\end{tikzcd}
\]
 \[
\begin{tikzcd}
 \arrow[r,"d_1"] & \mbox{End}(V_1) \oplus \mbox{Hom}(V_2,V_1)^{\oplus 2}\oplus \mbox{Hom}(V_2,W)
 \end{tikzcd}
\]
 \[
\begin{tikzcd}
 \arrow[r,"d_2"] &  \mbox{Hom}(V_2,V_1)
 \end{tikzcd}
\]
\caption{$\mathcal{C}(\mathcal{R})$} 
    \label{CR}
\end{figure}
with the differentials defined as
\squeezeup
\bea
&d_0(\alpha_1,\alpha_2)^t=([\alpha_1,A_1],[\alpha_2,A_2],\alpha_1,-J\alpha_1,[\alpha_2,B_2],\alpha_1f-f\alpha_2)^t\nonumber\\
&d_1(a_1,a_2,i,j,b_2,\phi)^t= \nonumber\\&([a_1,A_2]+[A_1,a_2]+Ij+iJ,A_1\phi+a_1f,A_2\phi+a_2f-fb_2-\phi B_2,jf+j\phi)^t\nonumber\\
&d_2(c_1,c_2,c_3,c_4)^t=c_1f+A_2c_2-c_2B_2-A_1c_3-Ic_4
\eea
Note that $H^1(\mathcal{C}(\mathcal{R}))$ parametrises the infinitesimal deformations of $\mathcal{R}$ and $H^2(\mathcal{C}(\mathcal{R}))$ the obstructions to the deformations. Moreover the conditions $H^0(\mathcal{C}(\mathcal{R}))=H^3(\mathcal{C}(\mathcal{R}))=0$ imply the stability of a framed representation.
 The equivariant virtual Euler characteristic of this determinant line bundle is arranged into a partition function of the quantum mechanical model.
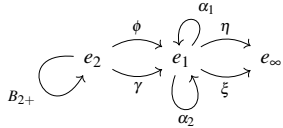
\begin{figure}
\begin{tikzpicture}
\begin{tikzcd}
\arrow[out=175,in=240,loop,swap,"B_{2+}"]
e_2 \arrow[r,bend left,"\phi"] 
\arrow[r,bend right,swap,"\gamma"] & e_1
 \arrow[out=50,in=100,loop,swap,"\alpha_1"]
 \arrow[out=250,in=300,loop,swap,"\alpha_2"]
  \arrow[r,bend left,"\eta"] \arrow[r,bend right,swap,"\xi"] & e_{\infty}
\end{tikzcd}
\end{tikzpicture}
\caption{enhanced ADHM quiver }
 \label{enquiver}
\end{figure}

The application of virtual equivariant localization requires the determination of fixed points of the torus action on the moduli space of the stable representations of the nested  ADHM  quivers.
The moduli space, $\mathcal{N} (r, n + d, d)$, for which we compute the  Euler characteristic,  the $\chi_y$ genus and the elliptic genus is described by the stable representations of an enhanced ADHM quiver of type $(n+d,d,r)$. Note that the parameter $r$ is the rank of the 5d gauge group and  $(n+d,d)$ are related to the number of  $D2,D2^{\prime}$ branes. The BPS counting function is  the Witten index of the supersymmetric quantum mechanics. The supersymmetric ground states are in one to one correspondence with cohomology classes in $\oplus_i H^{0,i}(\mathcal{N} (r, n_1, n_2))$. Note that in the limit of decoupling the surface operator  $\mathcal{N} (r, n + d, d)$ collapses to $\mathcal{M} (r, n )$ where $\mathcal{M} (r, n )$ is the suitably compactified and non-singular moduli space of instantons on $\mathbb{C}^2$. The later moduli space is isomorphic to the moduli space of rank $N$ torsion free sheaves with second Chern class $k$ on $\mathbb{P}^2$. So there exists a  morphism $q:\mathcal{N} (r, n + d, d) \to \mathcal{M} (r, n )$.  This morphism is equivariant with respect to the torus action. This makes it possible to classify the torus fixed loci in $\mathcal{N} (r, n + d, d)$.\\
In the following $(\mu,\nu)$ is a pair of nested Young diagrams satisfying the properties
\begin{itemize}
\item $\nu \subseteq \mu$
\item if $(i,j)\in \mu \setminus \nu, then\quad (i+1,j)\notin \mu$
\end{itemize}
Similarly if we denote an ordered sequence of Young diagrams by $\bf{\underline{\mu}}=\{\mu^1,...,\mu^r\}$ and $\bf{\underline{\nu}}=\{\nu^1,...,\nu^r\}$ then it is a nested sequence if it is pairwise $(\mu^a,\nu^a)$ nested. The numerical type of this sequence is given by $(|\underline{\mu|},|\underline{\nu}|)$. For nested sequence the defining algebraic inequalities are described as
\bea\label{eq:nestedcondition}
0\le c^a-e^a\le 1,\qquad 0\le \mu_i^a-\nu_i^a\le \nu_{i-1}^a-\nu_i^a
\eea
where $c^a,e^a$ denote the number of columns of $\mu^a$ and $\nu^a$ respectively and $a=1,...,r$ and $i\ge 0$. 
 The tangent space to the moduli space at a fixed point  $(\mu^a,\nu^a)$  of the torus $\bf{T}=\mathbb{C}^{\times}\times\mathbb{C}^{\times}\times(\mathbb{C}^{\times})^r$ is regarded as an element of the representation ring of the torus action and is given by the expression
\bea\label{fixedpointset}
T_{\underline{\mu},\underline{\nu}}\mathcal{N} (r, n + d, d)&=&T_{\underline{\nu}}\mathcal{M} (r, n )\nonumber\\&+&\sum_{a,b=1}^r\sum_{i=2}^{e^a+1}\sum_{j=1}^{\mu_j^b-\nu_j^b}R_a^{-1}R_bQ_1^{i-j}(Q_2^{\mu_i^a-\nu_j^b-s+1}-Q_2^{\nu^b_{i-1}-\nu_j^b-s+1})\nonumber\\&+&
\sum_{a,b=1}^r\sum_{j=1}^{c^b}\sum_{s=1}^{\mu_j^b-\nu_j^b}R_a^{-1}R_bQ_1^{-j+1}Q_2^{\mu_1^a-\nu_j^b-s+1}\nonumber\\
&=&\sum_{a,b=1}^rR_a^{-1}R_b\big(\sum_{(i,j)\in\nu^a}Q_1^{i-(\nu^b)^t_j}Q_2^{\nu_i^a-j+1}+\sum_{(i,j)\in\nu^b}Q_1^{(\nu^a)^t_j-i+1}Q_2^{j-\nu_i^b}  \big)\nonumber\\&+&\sum_{a,b=1}^r\sum_{i=2}^{e^a+1}\sum_{j=1}^{\mu_j^b-\nu_j^b}R_a^{-1}R_bQ_1^{i-j}(Q_2^{\mu_i^a-\nu_j^b-s+1}-Q_2^{\nu^b_{i-1}-\nu_j^b-s+1})\nonumber\\&+&
\sum_{a,b=1}^r\sum_{j=1}^{c^b}\sum_{s=1}^{\mu_j^b-\nu_j^b}R_a^{-1}R_bQ_1^{-j+1}Q_2^{\mu_1^a-\nu_j^b-s+1}\nonumber\\
\eea
where $R_1,...,R_r,Q_1,Q_2$ denote the one dimensional representations of $\bf{T}$ with their characters represented by $\rho_1,...,\rho_r,q_1,q_2$ respectively.\\
According to this formula the fixed point locus is a finite set of points in one to one correspondence  with pairs of nested sequences $(\underline{\mu},\underline{\nu})=(\mu^a,\nu^a)_{1\le a\le r}$ of length $r$. The type of Young diagrams is $(|\underline{\mu}|,|\underline{\nu}|)=(n+d,n)$.
\subsection{Holomorphic Euler characteristic, $\chi_y$ genus and elliptic genus}
Consider \cite{Fantechi_2010,Bonelli:2019het} a rank $d$ vector bundle $E$ on  some moduli space $X$. We can form the formal sums of the symmetric product $S_tE$ and the antisymmetric product $\Lambda_tE$ as
\bea
\Lambda_tE=\sum_{i=0}^d[\Lambda^iE]t^i,\quad S_tE=\sum_{i=0}[S^iE]t^i
\eea
The moduli space $X$ admits a virtual  cotangent bundle  $\Omega_X=(T_X)^{\vee}$ and the bundle of 
  n-forms $\Omega_X^{n}=\Lambda^n\Omega_X$.  On the scheme $X$ one can consider the perfect\footnote{perfect because the complex $E^{\bullet}$ has local isomorphism with a complex of vector bundles that is finite \cite{Fantechi_2010}.} obstruction theory $E^{\bullet}$ resolved to a complex of vector bundles $[E^{-1}\to E^0]$ and a virtual structure sheaf denoted by $\mathcal{O}_X^{vir}$.The virtual tangent bundle $T^{vir}_X$ is defined by the class $[E_0]-[E_1]$, where the complex $[E_0\to E_1]$  is dual to the complex $[E^{-1}\to E^0]$. The difference $\mbox{rank}(E_)-\mbox{rank}(E_1)$ defines what is called the virtual dimension of $X$. For a vector bundle $V$ given on $X$ and $[X]^{vir}$ the virtual fundamental class of $X$ as an element of the $(n-m)$-th Chow group of $X$ with rational coefficients, the virtual holomorphic Euler characteristic is defined by
  \bea
  \chi^{vir}(X,V)=\chi(X,V\otimes O_X^{vir})=\int_{[X]^{vir}}\mbox{ch}(V).\mbox{td}(T_X^{vir})
  \eea
   Then under suitable conditions  using the Riemann-Roch theorem the Hirzebruch $\chi_y$ genus is expressed as. See the appendix (\ref{virtuallocalization}) for a summary of the virtual localization.
  \bea
  \chi_{-y}(X,V)=\int_{[X]}\mbox{ch}(\Lambda_{-y}T_X)\mbox{ch}(V)\mbox{td}(T_X)=\int_{[X]}(\sum_{l=1}^de^{v_k})\prod_{i=1}^nx_i\frac{1-ye^{-x_i}}{1-e^{-x_i}}\prod_{j=1}^m\frac{1-e^{-u_i}}{u_j(1-ye^{-u_i})}\nonumber\\
  \eea
  where $x_1,...,x_n$ denotes the Chern roots of $E_0$, $u_1,...,u_m$ denote the Chern roots of $E_1$  and $v_1,..,v_r$ denote the Chern roots of the vector bundle $V$.
Note that for $y=0$, $\chi_{-y}^{vir}(X,V)$ reduces to the Euler characteristic
  \bea
  \chi(X,V)=\int_{[X]}(\sum_{l=1}^de^{v_k})\prod_{i=1}^n\frac{x_i}{1-e^{-x_i}}\prod_{j=1}^m\frac{1-e^{-u_i}}{u_j}\nonumber\\
  \eea
The moduli space $\mathcal{N} (r, n + d, d)$ is in general non-compact and the cohomology groups $H^{0,i}$ are not well defined. However due to the toric $\bf{T}=\mathbb{C}^{\times}\times \mathbb{C}^{\times}\times(\mathbb{C}^{\times})^r$ action on the moduli space, the Atiyah-Singer fixed point theorems can be applied to compute equivariant Euler character. The equivariant Euler character is an element of the quotient field of the representation ring of $\bf{T}$ and hence makes sense in the presence of non-compactness. The Euler character or the quiver partition function  $\chi_T(\mathcal{N} (r, n + d, d))$ can be generalised by coupling the quiver quantum mechanical system with the line bundles $\mathcal{L}_{(p_1,p_2)}$ parametrised by two integers $(p_1,p_2)\in\mathbb{Z}^2$. This coupling can be interpreted as the Chern Simons terms in the M-theory lift of the type IIA configuration. A generating function can thus be composed
\bea
Z^{4d,quiver}(q_1,q_2,\rho_a,Q)=\sum_{n\ge 0}\mbox{ch}_T\chi_T(\mathcal{N} (r, n + d, d),S\otimes \mathcal{L}_{(p_1,p_2)})
\eea
where $(q_1,q_2,\rho_1,...,\rho_r)$  denote the characters of the generators $(Q_1,Q_2,R_1,...,R_r,Q)$ as defined before.\\
Finally  the  generating function for the equivariant Euler characteristics  can be written as follows
\bea\label{eq:Z4quiver}
Z^{4d,quiver}(q_1,q_2,\rho_a,Q)&=& \sum_k Q^k \chi_T(k,q_1,q_2,\rho_a))\nonumber\\&=&\sum_kQ^k\sum_{|\nu|=k}\nonumber\\&\times&
\frac{1}{\prod_{(i,j)\in\nu^a}(1-q_1^{(\nu^{b})^t_j}q_2^{j-\nu_i^a-1})\prod_{(i,j)\in\nu^b}(1-q_1^{(\nu^{a})^t_j}q_2^{j-\nu_i^b-j})}\nonumber\\&\times&\sum_{\substack{(\mu,\nu)\\|\mu|=|\nu|+d}}
\Bigg[\prod_{i=1}^{c^a}\prod_{s=1}^{\mu^a_i-\nu^a_i}\rho_1 q_a^{1-i}q_2^{-\nu^a_i-s+1}\nonumber\\&\times&\frac{1}{\prod_{i=2}^{e^a+1}\prod_{j=1}^{c^b}\prod_{s=1}^{\mu^b_j-\nu_j^b}(1- q_1^{j-i}q_2^{\nu_j^b+s-\mu_i^a-1})}\nonumber\\ &\times&
\prod_{i=2}^{e^a+1}\prod_{j=1}^{c^b}\prod_{s=1}^{\mu^b_j-\nu_j^b}(1- q_1^{j-i}q_2^{\nu_j^b+s-\nu_{i-1}^a-1})
\nonumber\\&\times&
\frac{1}{\prod_{j=1}^{c^b}\prod_{s=1}^{\mu^b_j-\nu^b_j}(1- q_1^{j-1}q_2^{-\nu^b_j+s-\mu_1^a-1})}\Bigg]\nonumber\\&\equiv &\sum_kQ^k\sum_{|\nu|=k}\nonumber\\&\times&
\frac{1}{\prod_{(i,j)\in\nu^a}(1-q_1^{(\nu^{b})^t_j}q_2^{j-\nu_i^a-1})\prod_{(i,j)\in\nu^b}(1-q_1^{(\nu^{a})^t_j}q_2^{j-\nu_i^b-j})}\nonumber\\&\times&\sum_{\substack{(\mu,\nu)\\|\mu|=|\nu|+d}}W_{(\mu,\nu)}(q_1,q_2,\rho_a,y)
\eea
Here $W_{\nu,d}(q_1,q_2,\rho_a,y)\equiv\sum_{\substack{(\mu,\nu)\\|\mu|=|\nu|+d}}W_{(\mu,\nu)}(q_1,q_2,\rho_a,y)$ contains the information about the surface defect and the BPS states bound to it. \\
In \cite{Bruzzo:2010fk}  the refined open topological string partition function on the resolved conifold .i.e. the total space of $\mathcal{O}(-1)\oplus \mathcal{O}(-1) \to \mathbb{P}^1$, and the total space of the canonical bundle $\mathcal{O}(-2,-2)$ of $\mathbb{P}^1\times \mathbb{P}^1$ in the presence of topological D-branes was computed using the refined topological vertex formalism. Then these partition functions  are proved to be equal to the quiver partition function  (\ref{eq:Z4quiver}) 
 for $r=1$ and $r=2$ respectively.   More precisely one has to expand the open string partition function in terms of holonomy variables, say $x_i$, that parametrize D-brane boundary conditions, and  take the coefficient of $x_i^d$. Then this coefficient is to be compared with the gauge theory or quiver partition function.i.e. for $r=1$
\bea\label{eq:conjecture4d}
Z^{d,4d}_{quiver}(q_1,q_2,T)=Z^{ref}_{open,d}(q_1,q_2,T)
\eea
and for $r=2$
\bea
Z^{d,4d}_{quiver}(q_1,q_2,\rho_1,\rho_2,T)&=&Z^{ref}_{open,d}(q_1,q_2,\rho_{12},T)\nonumber\\
\eea
where the topological string partition function  $Z^{ref}_{open,d}$ is normalised  by dividing out by the gauge theory perturbative part. We elaborate on the refined topological vertex computation in the next section (\ref{section:chiy5d}) for the 'compactified' geometries. It is important to note that in computing the open string partition function  on the resolved conifold the defect brane was put on the un-preferred direction and the preferred direction was chosen to be the internal one.
Moreover in our case the lagrangian brane is put along an external leg. 
The open topological partition function contains both  perturbative and non-perturbative parts of gauge theory. In making the comparison (\ref{eq:conjecture4d}) one has to exclude the perturbative part. 
\section{Generating function of $\chi_y$ genus}\label{section:chiy5d}
Instanton partition functions for gauge theories in the presence of surface defect  in 4d, 5d and 6d are the generating functions of Euler characteristics, $\chi_y$-genera and elliptic genera of the  moduli space under consideration \cite{Li:2004ef}. 
For refined topological strings the  open string defect amplitude can be written in four ways depending on whether the topological brane extends along $\mathbb{R}^2_{\epsilon_1}$ or $\mathbb{R}^2_{\epsilon_2}$ and whether it is put along the external non-compact leg or the internal compact leg \cite{Kozcaz:2018ndf}. 
 \\
In the M-theory lift of the type IIA topological strings, the equivariant Euler characteristics gets lifted to the $\chi_y$ genus. It is a 5d defect gauge theory compactified on a circle $S^1$. The defect BPS states  in M-theory framework are related to  M2-brane BPS state counting. 
Consequently we write down the generating function for the $\chi_y$ genus  or $5d$ defect partition function as 
\bea
Z^{quiver}_d(q_1,q_2,\rho_a,y,Q)&=&\sum_kQ^k\sum_{|\nu|=k}\prod_{a,b=1}^r\nonumber\\&\times&
\frac{\prod_{(i,j)\in\nu^a}(1-y\rho_a\rho_b^{-1}q_1^{(\nu^{b})^t_j}q_2^{j-\nu_i^a-1})\prod_{(i,j)\in\nu^b}(1-y\rho_a\rho_b^{-1}q_1^{(\nu^{a})^t_j}q_2^{j-\nu_i^b-j})}{\prod_{(i,j)\in\nu^a}(1-\rho_a\rho_b^{-1}q_1^{(\nu^{b})^t_j}q_2^{j-\nu_i^a-1})\prod_{(i,j)\in\nu^b}(1-\rho_a\rho_b^{-1}q_1^{(\nu^{a})^t_j}q_2^{j-\nu_i^b-j})}\nonumber\\&\times&\sum_{\substack{(\underline{\mu},\underline{\nu})\\|\underline{\mu}|=|\underline{\nu}|+d}}
\Bigg[(\prod_{a=1}^r\prod_{i=1}^{c^a}\prod_{s=1}^{\mu^a_i-\nu^a_i}\rho_a q_1^{1-i}q_2^{-\nu^a_i-s+1})\nonumber\\&\times&\frac{\prod_{a,b=1}^r\prod_{i=2}^{e^a+1}\prod_{j=1}^{c^b}\prod_{s=1}^{\mu_j^b-\nu_j^b}(1-y\rho_a\rho_b^{-1} q_1^{j-i}q_2^{\nu^b_j+s-\mu^a_i-1})}{\prod_{a,b=1}^r\prod_{i=2}^{e^a+1}\prod_{j=1}^{c^b}\prod_{s=1}^{\mu_j^b-\nu_j^b}(1-\rho_a\rho_b^{-1} q_1^{j-i}q_2^{\nu^b_j+s-\mu^a_i-1})}\nonumber\\ &\times&
\frac{\prod_{a,b=1}^r\prod_{i=2}^{e^a+1}\prod_{j=1}^{c^b}\prod_{s=1}^{\mu^b_j-\nu^b_j}(1-\rho_a\rho_b^{-1} q_1^{j-i}q_2^{\nu^b_j+s-\nu^a_{i-1}-1})}{\prod_{a,b=1}^r\prod_{i=2}^{e^a+1}\prod_{j=1}^{c^b}\prod_{s=1}^{\mu^b_j-\nu^b_j}(1-y\rho_a\rho_b^{-1} q_1^{j-i}q_2^{\nu^b_j+s-\nu^a_{i-1}-1})}\nonumber\\&\times&
\frac{\prod_{a,b=1}^r\prod_{j=1}^{c^b}\prod_{s=1}^{\mu^b_j-\nu^b_j}(1-y \rho_a\rho_b^{-1}q_1^{j-1}q_2^{-\nu^b_j+s-\mu^a_1-1})}{\prod_{a,b=1}^r\prod_{j=1}^{c^b}\prod_{s=1}^{\mu^b_j-\nu^b_j}(1- \rho_a\rho_b^{-1}q_1^{j-1}q_2^{-\nu^b_j+s-\mu^a_1-1})}\Bigg]\nonumber\\
&\equiv &\sum_kQ^k\sum_{|\nu|=k}\nonumber\\&\times&
\frac{\prod_{(i,j)\in\nu^a}(1-yq_1^{(\nu^{b})^t_j}q_2^{j-\nu_i^a-1})\prod_{(i,j)\in\nu^b}(1-yq_1^{(\nu^{a})^t_j}q_2^{j-\nu_i^b-j})}{\prod_{(i,j)\in\nu^a}(1-q_1^{(\nu^{b})^t_j}q_2^{j-\nu_i^a-1})\prod_{(i,j)\in\nu^b}(1-q_1^{(\nu^{a})^t_j}q_2^{j-\nu_i^b-j})}\nonumber\\&\times&\sum_{\substack{(\underline{\mu},\underline{\nu})\\|\underline{\mu}|=|\underline{\nu}|+d}}W_{(\underline{\mu},\underline{\nu})}(q_1,q_2,\rho_a,y)
\eea
which for $r=1$ becomes
\bea\label{eq:Z5quiver}
Z^{5d,quiver}_d(q_1,q_2,\rho_1,Q)&=& \sum_k Q^k \chi_y(M,k,q_1,q_2,\rho_1))\nonumber\\&=&\sum_kQ^k\sum_{|\nu|=k}\nonumber\\&\times&
\frac{\prod_{(i,j)\in\nu}(1-yq_1^{(\nu)^t_j}q_2^{j-\nu_i-1})\prod_{(i,j)\in\nu}(1-yq_1^{(\nu)^t_j}q_2^{j-\nu_i-j})}{\prod_{(i,j)\in\nu}(1-q_1^{(\nu)^t_j}q_2^{j-\nu_i-1})\prod_{(i,j)\in\nu}(1-q_1^{(\nu)^t_j}q_2^{j-\nu_i-j})}\nonumber\\&\times&\sum_{\substack{(\mu,\nu)\\|\mu|=|\nu|+d}}
\Bigg[\prod_{i=1}^{c}\prod_{s=1}^{\mu_i-\nu_i}\rho_1 q_1^{1-i}q_2^{-\nu_i-s+1}\nonumber\\&\times&\frac{\prod_{i=2}^{e+1}\prod_{j=1}^{c}\prod_{s=1}^{\mu_j-\nu_j}(1-y q_1^{j-i}q_2^{\nu_j+s-\mu_i-1})}{\prod_{i=2}^{e+1}\prod_{j=1}^{c}\prod_{s=1}^{\mu_j-\nu_j}(1- q_1^{j-i}q_2^{\nu_j+s-\mu_i-1})}\nonumber\\ &\times&
\frac{\prod_{i=2}^{e+1}\prod_{j=1}^{c}\prod_{s=1}^{\mu_j-\nu_j}(1- q_1^{j-i}q_2^{\nu_j+s-\nu_{i-1}-1})}{\prod_{i=2}^{e+1}\prod_{j=1}^{c}\prod_{s=1}^{\mu_j-\nu_j}(1- yq_1^{j-i}q_2^{\nu_j+s-\nu_{i-1}-1})}\nonumber\\&\times&
\frac{\prod_{j=1}^{c}\prod_{s=1}^{\mu_j-\nu_j}(1-y q_1^{j-1}q_2^{-\nu_j+s-\mu_1-1})}{\prod_{j=1}^{c}\prod_{s=1}^{\mu_j-\nu_j}(1- q_1^{j-1}q_2^{-\nu_j+s-\mu_1-1})}\Bigg]\nonumber\\&\equiv &\sum_kQ^k\sum_{|\nu|=k}\nonumber\\&\times&
\frac{\prod_{(i,j)\in\nu^a}(1-yq_1^{(\nu)^t_j}q_2^{j-\nu_i-1})\prod_{(i,j)\in\nu}(1-yq_1^{(\nu)^t_j}q_2^{j-\nu_i-j})}{\prod_{(i,j)\in\nu^a}(1-q_1^{(\nu)^t_j}q_2^{j-\nu_i-1})\prod_{(i,j)\in\nu}(1-q_1^{(\nu)^t_j}q_2^{j-\nu_i-j})}\nonumber\\&\times&\sum_{\substack{(\mu,\nu)\\|\mu|=|\nu|+d}}W_{(\mu,\nu)}(q_1,q_2,\rho_1,y)
\eea
Here $W_{\nu,d}(q_1,q_2,\rho_1,y)\equiv\sum_{\substack{(\mu,\nu)\\|\mu|=|\nu|+d}}W_{(\mu,\nu)}(q_1,q_2,\rho_1,y)$ contains the information about  the surface defects and the BPS states bound to it.
\subsection{Open string/defect brane partition function $\mathcal{O}(-1)\oplus \mathcal{O}(-1) \to \mathbb{P}^1$: gauging the two external legs }
\begin{figure}
\begin{tikzpicture}
\draw[gray, thick] (2,4) -- (3,4);
\draw[red, thick,dashed] (2.4,4)-- (3,4.9);
\draw[gray, thick] (2,4) -- (2,5);
\draw[blue, thick] (1.6,4.5) -- (2.4,4.5);
\draw[gray, thick] (0,2) -- (2,4);
\draw[green, thick]  (0.5,3.4) -- (1.3,2.4);
\node [right] at (1,3) {$Q_2$};
\draw[gray, thick] (0,2) -- (0,1);
\draw[blue, thick] (-0.5,1.5) -- (0.4,1.5);
\draw[gray, thick] (0,2) -- (-1,2);
\node [right] at (0.0,1.7){$Q_1$};
\end{tikzpicture}
\caption{partially compactified toric diagram of resolved conifold with a Lagrangian brane}
 \label{compression}
\end{figure}
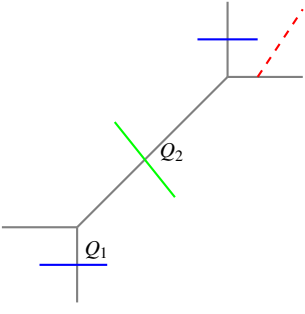
The resolved conifold i.e. the total space of the bundle $\mathcal{O}(-1)\oplus \mathcal{O}(-1)\to \mathbb{P}^1$ can be parametrised by the pair of coordinates $(y_1,y_2)$ and $(y_3,y_4)$. These pairs of coordinates define two line bundles on $\mathbb{P}^1$ as follows
\bea\label{eq:rescon}
y_1= \zeta y_2,\quad y_3=\zeta^{-1}y_4
\eea
for $\zeta\in \mathbb{P}^1$. The  partial compactification of the corresponding web diagram imposes periodic boundary conditions on the external legs.\\
Note that in the toric diagram (\ref{compression}) two blue lines indicate the identification of vertical edges, the green line intersects the preferred direction and the red dotted line denotes the lagrangian brane.
For our purposes it is more useful to choose the internal line as our preferred direction. With the following definitions of 
the framing factor $f_{\mu}(t,q)$ and the refined topological vertex  $C_{\lambda\mu\nu}(t,q)$ given by \cite{Bruzzo:2010fk}
\bea\label{eq:RTV2}
f_{\mu}(t,q)&=&(-1)^{|\mu|}t^{||\mu^t||/2-|\mu|/2}q^{-||\mu||^2/2+|\mu|/2}\nonumber\\
C_{\lambda\mu\nu}(t,q)&=&(\frac{t}{q})^{\frac{||\mu||^2}{2}}q^{\frac{\kappa(\mu)+||\nu||^2}{2}}\tilde{Z}_{\nu}(t,q)\nonumber\\&\times&\sum_{\eta}(\frac{q}{t})^{\frac{|\eta|+|\lambda|-|\mu|}{2}}s_{\lambda^t/\eta}(t^{-\rho}q^{-\nu})s_{\mu/\eta}(t^{-\nu^t}q^{-\rho})\nonumber\\
\tilde{Z}_{\nu}(t,q)&=&\prod_{i=1}^{l(\nu)}\prod_{j=1}^{\lambda_i}(1-q^{\lambda_i-j}t^{\lambda_j^t-i+1})^{-1}\nonumber\\
\rho&=&\{-\frac{1}{2},-\frac{3}{2},-\frac{5}{2},...\},\quad t^{-\rho}q^{-\nu}=\{t^{\frac{1}{2}}q^{-\lambda_1},t^{\frac{3}{2}}q^{-\lambda_2},t^{\frac{5}{2}}q^{-\lambda_3},... \}
\eea
 the refined amplitude can be written in the following form
\bea
Z^{ref}_{open}&=&\sum s_{\lambda}(x)\mathcal{Z}_{\lambda}(Q_1,Q_2,t,q)\nonumber\\&=&\sum_{\nu} s_{\lambda}(x)\sum_{\mu,\lambda}(-Q_1)^{|\nu |}(-Q_2)^{|\mu |}C_{\emptyset\mu\nu}(t,q)C_{\lambda\mu^{t}\nu^t}(q,t)\nonumber\\&=&
\sum_{\nu}(-Q_1)^{|\nu |}(\frac{q}{t})^{\frac{||\nu||^2-||\nu^t||^2}{2}}P_{\nu^t}(t^{-\rho};q,t)P_{\nu}(q^{-\rho};t,q)
\prod_{i,j}(1+(\sqrt{t}x_j) t^{-\nu_i^t}q^{i-1})\nonumber\\&\times&\prod_{k,l}(1-Q_2 t^{-\nu^t_k-\rho_l}q^{-\rho_k-\nu_l})\prod_{i,j}(1+Q_2\sqrt{\frac{q}{t}}(\sqrt{t}x_j) t^{-\nu_i^t}q^{i-1})^{-1}\nonumber\\
\eea
where
\bea
 P_{\nu^t}(t^{-\rho};q,t)&=&t^{\frac{||\nu||^2}{2}}\tilde{Z}_{\nu}(t,q)\nonumber\\
 &=&t^{\frac{||\nu||^2}{2}}\prod_{(i,j)\in\nu}(1-t^{a(i,j)+1}q^{l(i,j)})^{-1}
 \nonumber\\
 &=&t^{\frac{||\nu||^2}{2}}\prod_{(i,j)\in\nu}(1-t^{v_j^t-i+1}q^{\nu_i-j})^{-1}
 \eea
After dividing by the gauge theory perturbative  factor $\prod_{k,l}(1-Q_2 t^{-\rho_l}q^{-\rho_k})$ 
and using the  following identity \cite{Iqbal:2007ii}
\bea
\frac{\prod_{k,l}(1-Q_2 t^{-\mu^t_k-\rho_l}q^{-\rho_k-\mu_l})}{\prod_{k,l}(1-Q_2 t^{-\rho_l}q^{-\rho_k})}=\prod_{s\in\mu}(1-Q_2 t^{-a(s)_k-\frac{1}{2}}q^{-l(s)_k-\frac{1}{2}})(1-Q_2 t^{a(s)_k+\frac{1}{2}}q^{l(s)_k+\frac{1}{2}})
\eea
where $a(i,j)=\nu_j^t-i$, $l(i,j)=\nu_i-j$ denote the arm length and leg length of a box at the position $(i,j)$ in the Young diagram,
we get in the redefined variables $y_j=x_j\sqrt{t}$
\bea
\tilde{Z}^{ref}_{open}&=&
\sum_{\nu}(-Q_1)^{|\nu |}(\frac{q}{t})^{\frac{||\nu||^2-||\nu^t||^2}{2}}P_{\nu^t}(t^{-\rho};q,t)P_{\nu}(q^{-\rho};t,q)
\nonumber\\&\times&\prod_{s\in\nu}(1-Q_2 t^{-a(s)_k-\frac{1}{2}}q^{-l(s)_k-\frac{1}{2}})(1-Q_2 t^{a(s)_k+\frac{1}{2}}q^{l(s)_k+\frac{1}{2}})\nonumber\\&\times&\prod_{i,j}(1+y_j t^{-\nu_i^t}q^{i-1})\prod_{m,n}(1+Q_2\sqrt{\frac{q}{t}}y_j t^{-\nu_i^t}q^{i-1})^{-1}\nonumber\\
\eea
Now to extract the contribution of the surface operator we proceed as suggested in \cite{Bruzzo:2010fk}.
The right hand side of the last equation can be expanded in the basis of symmetric functions, in particular in the basis of monomial symmetric functions $m_{\nu}(x_i)$ \cite{MacDonald:2019755}. By definition, $m_{n,0,0,...}(\mbox{\bf{x}})=x_1^n+x_2^n+...$
for any positive integer  $n$. Then we will denote by $\tilde{Z}^{ref}_{open,d}$ the coefficient of $m_{d,0,0,...}(\bf{x})$ in the expansion. \\
Next note that 
\bea\label{eq:symm1}
\mbox{ln}\big(\prod_{i,j}(1+y_j t^{-\nu_i^t}q^{i-1})\big)=\sum_{i=1}^{\infty}\sum_{l=1}^{\infty}\frac{(-1)^{l-1}}{l}(\sum_{k=1}^{\infty}t^{-k\nu_i^t}q^{k(i-1)}e_k(\mbox{\bf{y}}))^l
\eea
where $e_k(\bf{y})$, $k\in\mathbb{Z}_{\ge 0}$ denotes the degree $k$ elementary symmetric function in the variables $(y_1,y_2,...,)$. Recall the fact that for a partition $\lambda$ and its conjugate partition $\lambda^{\prime}$ we have the expansion \cite{MacDonald:2019755}
\bea\label{eq:me}
e_{\lambda^{\prime}}=m_{\lambda}+\sum_{\mu}a_{\lambda\mu}m_{\mu}
\eea
where the summation over $\mu$ is such that $\mu<\lambda$ and $a_{\lambda\mu}$ are non-negative integers.
Using the identity (\ref{eq:me}) it is easy to see that  the coefficient of $m_{d,0,0,...}(\mbox{\bf{x}})=x_1^d+x_2^d+...$ in the expansion (\ref{eq:symm1}) can be obtained by restricting to $k=1$
\bea
e^{\sum_{i=1}^{\infty}\sum_{l=1}^{\infty}\frac{(-1)^{l-1}}{l}(t^{-\nu_i^t}q^{(i-1)}e_1(\mbox{\bf{y}}))^l}=e^{\sum_{l=1}^{\infty}\frac{(-1)^{l-1}}{l} \sum_{i=1}^{\infty}(t^{-l\nu_i^t}q^{l(i-1)}e_1(\mbox{\bf{y}}))^l}
\eea
Defining the quantity $F_{\nu}$ by
\bea
F_{\nu}(q,t)=\sum_{i=1}^{\infty}q^{i-1}t^{-\nu_i^t}
\eea
we can write
\bea
e^{\sum_{i=1}^{\infty}\sum_{l=1}^{\infty}\frac{(-1)^{l-1}}{l}(t^{-\nu_i^t}q^{(i-1)}e_1(\mbox{\bf{y}}))^l}=e^{\sum_{l=1}^{\infty}\frac{(-1)^{l-1}}{l} F_{\nu}(q^l,t^l)e_1(\mbox{\bf{y}}))^l}
\eea
From the last expression we find the coefficient of $m_{d,0,0,,,,}$ as
\bea
\frac{1}{d!}\sum_{\eta=(1^{d_1},2^{d_2},...)}\frac{d!}{\prod_{k=1}^d d_k!}\prod_{k=1}^d\big(\frac{(-1)^{k-1}}{k}F_{\nu}(q,t) \big)
\eea
where $\eta=(1^{d_1},2^{d_2},...)$ denotes the set of all partitions of $d$.\\
Following the same procedure  we find $\tilde{Z}^{ref}_{open,d}$ as 
\bea
\tilde{Z}^{ref}_{open,d}&=&
\sum_{\mu}(-Q_1)^{|\mu |}(\frac{q}{t})^{\frac{||\mu||^2-||\mu^t||^2}{2}}P_{\mu^t}(t^{-\rho};q,t)P_{\mu}(q^{-\rho};t,q)
\nonumber\\&\times&\prod_{s\in\mu}(1-Q_2 t^{-a(s)_k-\frac{1}{2}}q^{-l(s)_k-\frac{1}{2}})(1-Q_2 t^{a(s)_k+\frac{1}{2}}q^{l(s)_k+\frac{1}{2}})\nonumber\\&\times&
\sum_{\eta_1=(1^{d_1},2^{d_2},...,d^{d_d})}\frac{(-1)^{d-\sum_{k=1}^dd_k}}{\prod_{k=1}^d(d_k!k^{d_k})}\prod_{k=1}^d F_{\mu}(t^k,q^k,Q_2)^{d_k}
\eea
where
\bea
F_{\nu}(t,q,Q_1)&=&\sum_{i=1}q^{i-1}t^{-\nu_i^t}-Q_2\sqrt{\frac{q}{t}}\sum_{i=1}q^{i-1}t^{-\nu_i^t}
\nonumber\\
\eea
making  a change of variables
\bea
q=q_2^{-1},\quad t=q_1
\eea
\bea
F_{\nu}(t,q,Q_2)
&=&\bigg((\sum_{i=1}q_2^{1-i}q_1^{-\nu^t_i}-Q_2\sqrt{\frac{q}{t}}\sum_{i=1}q_2^{1-i}q_1^{-\nu_i^t})\bigg)\nonumber\\
\eea
The 5d generalisation of  (\ref{eq:conjecture4d}) states
\bea
Z^{5d,quiver}_d=\tilde{Z}^{ref}_{open,d}
\eea
or explicitly
\bea\label{eq:5dconj}
&\sum_kQ^k\sum_{|\nu|=k}
\frac{\prod_{(i,j)\in\nu}(1-yq_1^{(\nu)^t_j}q_2^{j-\nu_i-1})\prod_{(i,j)\in\nu}(1-yq_1^{(\nu)^t_j}q_2^{j-\nu_i-j})}{\prod_{(i,j)\in\nu}(1-q_1^{(\nu)^t_j}q_2^{j-\nu_i-1})\prod_{(i,j)\in\nu}(1-q_1^{(\nu)^t_j}q_2^{j-\nu_i-j})}\nonumber\\&\times\sum_{\substack{(\mu,\nu)\\|\mu|=|\nu|+d}}W_{(\mu,\nu)}(q_1,q_2,\rho_1,y)=_{parameter-identification}\nonumber\\&\times
\sum_{\mu}(-Q_1)^{|\mu |}(\frac{q}{t})^{\frac{||\mu||^2-||\mu^t||^2}{2}}P_{\mu^t}(t^{-\rho};q,t)P_{\mu}(q^{-\rho};t,q)
\nonumber\\&\times\prod_{s\in\mu}(1-Q_2 t^{-a(s)_k-\frac{1}{2}}q^{-l(s)_k-\frac{1}{2}})(1-Q_2 t^{a(s)_k+\frac{1}{2}}q^{l(s)_k+\frac{1}{2}})\nonumber\\&\times
\sum_{\eta_1}\frac{(-1)^{d-\sum_{k=1}^dd_k}}{\prod_{k=1}^d(d_k!k^{d_k})}\prod_{k=1}^d F_{\mu}(t^k,q^k,Q_2)^{d_k}
\eea
where $parameter-identification$ is described in the section \ref{DTKT}.
We have to divide by $Z_{pert}$ since the refined topological vertex technique computes both perturbative and non-perturbative contributions from the gauge theory point of view, whereas $W_{\nu,d}(q_1,q_2,y)$ only describes non-perturbative contributions.
\subsection{special case of the conjecture}
\subsection*{$\nu=\emptyset$}
We know that in the absence of a Lagrangian brane we have to set
\bea
\nu=\emptyset, 
\eea 
 and the conjecture (\ref{eq:5dconj}) reduces to the special case
\bea\label{eq:specialchiy}
&\sum_kQ^k\sum_{|\nu|=k}
\frac{\prod_{(i,j)\in\nu}(1-yq_1^{(\nu)^t_j}q_2^{j-\nu_i-1})\prod_{(i,j)\in\nu}(1-yq_1^{(\nu)^t_j}q_2^{j-\nu_i-j})}{\prod_{(i,j)\in\nu}(1-q_1^{(\nu)^t_j}q_2^{j-\nu_i-1})\prod_{(i,j)\in\nu}(1-q_1^{(\nu)^t_j}q_2^{j-\nu_i-j})}=_{parameter-identification}\nonumber\\&\times
\sum_{\mu}(-Q_1)^{|\mu |}(\frac{q}{t})^{\frac{||\mu||^2-||\mu^t||^2}{2}}P_{\mu^t}(t^{-\rho};q,t)P_{\mu}(q^{-\rho};t,q)
\nonumber\\&\times\prod_{s\in\mu}(1-Q_2 t^{-a(s)_k-\frac{1}{2}}q^{-l(s)_k-\frac{1}{2}})(1-Q_2 t^{a(s)_k+\frac{1}{2}}q^{l(s)_k+\frac{1}{2}})
\eea
Using the following definition of the Macdonald symmetric function $P_{\nu}(\mbox{\bf{x}};q,t)$ 
\bea
P_{\nu^t}(t^{-\rho};q,t)=t^{\frac{||\nu||^2}{2}}\prod_{s\in\nu}(1-t^{a(s)+1}q^{l(s)})=t^{\frac{||\nu||^2}{2}}\prod_{s\in\nu}(1-t^{l(s)+1}q^{a(s)})
\eea
the identity (\ref{eq:specialchiy})
 is satisfied \cite{Iqbal:2007ii} by taking 
\bea
t=q_1,\quad q=q_2^{-1},\quad Q_2=y\sqrt{\frac{t}{q}},\quad Q_1=\sqrt{\frac{t}{q}}Q
\eea
\subsection*{$\nu\ne \emptyset$}
 From the 2d quiver quantum mechanical point of view  $y$ is the mass deformation parameter \cite{Hollowood:2003gr} and does not depend on $\nu$, the representation of the lagrangian branes. Thus turning on $\nu$ from $\nu=\emptyset$ to $\nu \ne \emptyset$ does not change the dependence on the mass parameter $y$. Note that for $y=1$ and $q_1=q_2^{-1}:=q$
 \bea
Z^{5d,quiver}_d(q,\rho_a,Q)&=& \sum_k Q^k \chi_y(M,k,q,\rho_a))\nonumber\\&\equiv &\sum_kQ^k\sum_{|\nu|=k}\sum_{\substack{(\underline{\mu},\underline{\nu})\\|\mu|=|\nu|+d}}(1)
\eea
where the summation  $\sum_{\substack{(\underline{\mu},\underline{\nu})\\|\mu|=|\nu|+d}}(1)$ counts the number of nested partitions $(\mu_a,\nu_a)$ for a given $\underline{\nu}$ such that \\
(a) no two points in the complements $\mu_a/\nu_a$ are in the same row\\ or in other words $(\mu_a,\nu_a)$ and their respective number of columns $(c^a,e^a)$ satisfy the following constraints\\
(b)
$0\le c^a-e^a\le 1,\qquad 0\le \mu_i^a-\nu_i^a\le \nu_{i-1}^a-\nu_i^a$ \quad
for $1 \le a\le r$ and $i>0$,
\subsection{Analytic Continuation}\label{anylcont}
Although the contribution due to the topological brane does not seem to be polynomial in $y$, it is easy to see that after analytically continuing $q_2\to q_2^{-1}$, $W_{\nu,d}(q_1,q_2,y,\rho_a)$ can be written in the following form
\bea\label{eq:defect}
&W&_{\nu,d}(q_1,q_2,y,\rho_a)=\sum_{\substack{(\underline{\mu},\underline{\nu})\\|\underline{\mu}|=|\underline{\nu}|+d}}
\Bigg[\prod_{i=1}^{c^a}\prod_{s=1}^{\mu^a_i-\nu^a_i}\rho_1 q_1^{1-i}q_2^{-\nu^a_i-s+1}\prod_{a,b=1}^r\nonumber\\&\times&\bigg(\frac{\prod_{i=2}^{e^a+1}\prod_{j=1}^{c^b}\prod_{s=1}^{\mu^b_j-\nu^b_j}(1-y\rho_a\rho_b^{-1} q_1^{j-i}q_2^{\nu_j^b+s-\mu^a_i-1})}{\prod_{i=2}^{e^a+1}\prod_{j=1}^{c^b}\prod_{s=1}^{\mu^b_j-\nu^b_j}(1-\rho_a\rho_b^{-1} q_1^{j-i}q_2^{\nu_j^b+s-\mu^a_i-1})}
\frac{\prod_{j=1}^{c^b}\prod_{s=1}^{\mu^b_j-\nu^b_j}(1-\rho_a\rho_b^{-1}y q_1^{j-1}q_2^{-\nu^b_j+s-\mu^a_1-1})}{\prod_{j=1}^{c^b}\prod_{s=1}^{\mu^b_j-\nu^b_j}(1- \rho_a\rho_b^{-1}q_1^{j-1}q_2^{-\nu^b_j+s-\mu^a_1-1})}\nonumber\\ &\times&
\frac{\prod_{i=2}^{e^a+1}\prod_{j=1}^{c^b}\prod_{r=0}(1-y \rho_a\rho_b^{-1}q_1^{j-i}q_2^{\nu_j^b-\nu_{i-1}^a-r-1})(1-y \rho_a\rho_b^{-1}q_1^{j-i}q_2^{\mu_j^b-\nu_{i-1}^a+r})}{\prod_{i=2}^{e^a+1}\prod_{j=1}^{c^b}\prod_{r=0}(1- \rho_a\rho_b^{-1}q_1^{j-i}q_2^{\nu_j^b-\nu_{i-1}^a-r-1})(1- \rho_a\rho_b^{-1}q_1^{j-i}q_2^{\mu_j^b-\nu_{i-1}^a+r})}\bigg)
\Bigg]\nonumber\\&=&
W_{\nu,d}(q_1,q_2,\rho_a)+\mathcal{O}(y^d)
\eea
Noting that
\bea
\prod_{i=1}^{c^a}\prod_{s=1}^{\mu^a_i-\nu^a_i}\rho_1 q_1^{1-i}q_2^{-\nu^a_i-s+1}
=\sqrt{q_1q_2}^{d+(||\underline{\nu}||^2-||\underline{\mu}^2||)}
\eea
\bea
&W&_{\nu,d}(q_1,q_2,y,\rho_a)=\sum_{\substack{(\underline{\mu},\underline{\nu})\\|\underline{\mu}|=|\underline{\nu}|+d}}
\Bigg[\sqrt{q_1q_2}^{d+(||\underline{\nu}||^2-||\underline{\mu}^2||)}\prod_{a,b=1}^r\nonumber\\&\times&\bigg(\frac{\prod_{i=2}^{e^a+1}\prod_{j=1}^{c^b}\prod_{s=1}^{\mu^b_j-\nu^b_j}(1-y\rho_a\rho_b^{-1} q_1^{j-i}q_2^{\nu_j^b+s-\mu^a_i-1})}{\prod_{i=2}^{e^a+1}\prod_{j=1}^{c^b}\prod_{s=1}^{\mu^b_j-\nu^b_j}(1-\rho_a\rho_b^{-1} q_1^{j-i}q_2^{\nu_j^b+s-\mu^a_i-1})}
\frac{\prod_{j=1}^{c^b}\prod_{s=1}^{\mu^b_j-\nu^b_j}(1-\rho_a\rho_b^{-1}y q_1^{j-1}q_2^{-\nu^b_j+s-\mu^a_1-1})}{\prod_{j=1}^{c^b}\prod_{s=1}^{\mu^b_j-\nu^b_j}(1- \rho_a\rho_b^{-1}q_1^{j-1}q_2^{-\nu^b_j+s-\mu^a_1-1})}\nonumber\\ &\times&
\frac{\prod_{i=2}^{e^a+1}\prod_{j=1}^{c^b}\prod_{r=0}(1-y \rho_a\rho_b^{-1}q_1^{j-i}q_2^{\nu_j^b-\nu_{i-1}^a-r-1})(1-y \rho_a\rho_b^{-1}q_1^{j-i}q_2^{\mu_j^b-\nu_{i-1}^a+r})}{\prod_{i=2}^{e^a+1}\prod_{j=1}^{c^b}\prod_{r=0}(1- \rho_a\rho_b^{-1}q_1^{j-i}q_2^{\nu_j^b-\nu_{i-1}^a-r-1})(1- \rho_a\rho_b^{-1}q_1^{j-i}q_2^{\mu_j^b-\nu_{i-1}^a+r})}\bigg)
\Bigg]
\eea
\subsection{Generating function of $\chi_y$ genus: $r=2$
}\label{sectionr2}
For this case the partially compactified toric diagram of the  total space of the bundle $\mathcal{O}(-2,-2)$  of $\mathbb{P}^1\times\mathbb{P}^1$ is given in figure \ref{F0}. Note that the edges which are identified are parallel.
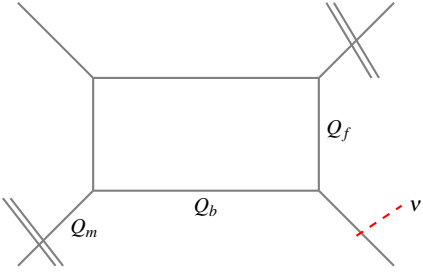
\begin{figure}
\begin{tikzpicture}
\draw[gray, thick] (2,4) -- (5,4);
\draw[gray, thick] (5,4) -- (6,5);
\draw[gray, thick] (5.8,4) -- (5.2,5);
\draw[gray, thick] (5.7,4) -- (5.1,5);
\draw[gray, thick] (5,4) -- (5,2.5);
\draw[gray, thick] (5,2.5) -- (6,1.5);
\draw[red, thick,dashed] (5.5,1.9)-- (6.1,2.3);
\node [right] at (6.1,2.3) {$\nu$};
\draw[gray, thick] (5,2.5) -- (2,2.5);
\draw[gray, thick] (2,2.5) -- (1,1.5);
\draw[gray, thick] (1.5,1.5) -- (0.8,2.4);
\draw[gray, thick] (1.6,1.5) -- (0.9,2.4);
\node [right] at (1.6,2) {$Q_m$};
\draw[gray, thick](2,2.5) -- (2,4);
\draw[gray, thick](2,4) -- (1,5);
\node [right] at (5,3.3) {$Q_f$};
\node [below] at (3.5,2.5) {$Q_b$};
\end{tikzpicture}
\caption{partially compactified toric diagram of  total space of the bundle $\mathcal{O}(-2,-2)$  of $\mathbb{P}^1\times\mathbb{P}^1$}
    \label{F0}
\end{figure}
\begin{figure}
\begin{tikzpicture}
\draw[black, thick] (6,4) -- (7,4)node[above]{$\tilde{Q}_1^{\prime}$};
\draw[black, thick] (6,3) -- (7,3)node[below]{$\tilde{Q}_2^{\prime}$};
\draw[black, thick] (6,4) -- (6,3)node[above]{$Q_1^{\prime}$};
\draw[black, thick] (6,4) -- (5.5,4.5)node[right]{};
\draw[black, thick] (6,3) -- (5.5,2.5)node[right]{$Q_{\rho}^{\prime}$};
\draw[black, thick] (5.5,2.5) -- (4.5,2.5);
\draw[black, thick] (5.5,1.5) -- (4.5,1.5);
\draw[black, thick] (5.5,2.5) -- (5.5,1.5)node[above]{$Q_2^{\prime}$};
\draw[black, thick] (5.5,1.5) -- (6.5,0.6)node[right]{$\nu^t$};
\draw[red, thick,dashed] (6.1,1) -- (6.7,1.3);
\draw[blue, thick] (5,1.3) -- (5,1.7);
\draw[blue, thick] (5.1,1.3) -- (5.1,1.7);
\draw[blue, thick] (6.5,2.8) -- (6.5,3.2);
\draw[blue, thick] (6.6,2.8) -- (6.6,3.2);
\draw[green, thick] (6.5,3.8) -- (6.5,4.2);
\draw[green, thick] (6.6,3.8) -- (6.6,4.2);
\draw[green, thick] (5,2.3) -- (5,2.7);
\draw[green, thick] (5.1,2.3) -- (5.1,2.7);
\end{tikzpicture}
\caption{equivalent to toric diagram in figure \ref{F0}}
\label{equiv}
\end{figure}
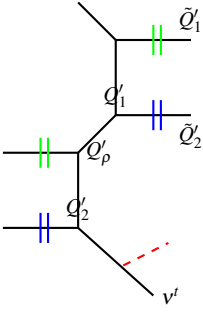
In our computation we choose the horizontal direction as the preferred one and hence the refined topological vertex can be used.  An important ingredient in the refined vertex computation is the choice of preferred direction, which should be common to all of the vertices in the web-digram. However if we had chosen the direction along which the lagrangian brane is present as the preferred direction, it is not shared by two of the vertices in the web diagram (\ref{F0}). This requires the introduction of a new refined topological vertex \cite{Iqbal:2012mt}. \\ 
To avoid the subtleties of  the new refined topological vertex  we can instead choose to compute refined partition function of the flopped geometry \cite{AIqbalsms,Bastian:2018fba}. Although for the horizontal preferred direction this is not necessary, we use it to illustrate the procedure. The flopped geometry is obtained from the original geometry by moving in the moduli space  defined by the extended K\"ahler cone of the CY 3-fold under consideration. 
For instance the geometry defined in (\ref{eq:rescon}),has its flopped version defined as
\bea\label{eq:resconflop}
y_1= \tilde{\zeta} y_4,\quad y_2=\tilde{\zeta}^{-1}y_3
\eea
for $\tilde{\zeta}\in \mathbb{P}^1$.
 The toric Calabi Yau manifolds have the important property that they can be converted to a strip geometry form after appropriate number of blow ups and flop operations \cite{Iqbal:2004ne}. \\ The flopped geometry contains as building blocks partially compactified $\mathcal{O}(-1)\oplus \mathcal{O}(-1)\to\mathbb{P}^1$s suitably glued together. 
To perform flop transition it is useful to draw the toric diagram in an equivalent way, figure \ref{equiv}, which makes the appearance of the  building blocks $\mathcal{O}(-1)\oplus \mathcal{O}(-1)\to\mathbb{P}^1$, $\mathcal{O}(-2)\oplus \mathcal{O}(0)\to\mathbb{P}^1$ and $\mathcal{O}(0)\oplus \mathcal{O}(-2)\to\mathbb{P}^1$  manifest.
Finally performing the flop on $\mathbb{P}^1$ whose normal bundle is $\mathcal{O}(-1)\oplus \mathcal{O}(-1)$, results in the web diagram given in  figure \ref{flopped}. 
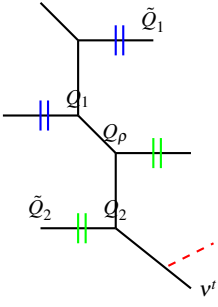
\begin{figure}
\begin{tikzpicture}
\draw[black, thick] (10,4) -- (11,4)node[above]{$\tilde{Q}_1^{}$};
\draw[black, thick] (10,3) -- (9,3)node[below]{$$};
\draw[black, thick] (10,4) -- (10,3)node[above]{$Q_1$};
\draw[black, thick] (10,4) -- (9.5,4.5)node[right]{$$};
\draw[black, thick] (10,3) -- (10.5,2.5)node[above]{$Q_{\rho}$};
\draw[black, thick] (10.5,2.5) -- (11.5,2.5);
\draw[black, thick] (10.5,2.5) -- (10.5,1.5)node[above]{$Q_2^{}$};
\draw[black, thick] (10.5,1.5) -- (9.5,1.5)node[above]{$\tilde{Q}_2^{}$};
\draw[black, thick] (10.5,1.5) -- (11.5,0.7)node[right]{$\nu^t$};
\draw[red, thick,dashed] (11.2,1) -- (11.8,1.3);
\draw[green, thick] (10,1.3) -- (10,1.7);
\draw[green, thick] (10.1,1.3) -- (10.1,1.7);
\draw[green, thick] (11,2.3) -- (11,2.7);
\draw[green, thick] (11.1,2.3) -- (11.1,2.7);
\draw[blue, thick] (10.5,3.8) -- (10.5,4.2);
\draw[blue, thick] (10.6,3.8) -- (10.6,4.2);
\draw[blue, thick] (9.5,2.8) -- (9.5,3.2);
\draw[blue, thick] (9.6,2.8) -- (9.6,3.2);
\end{tikzpicture}
\caption{geometry after the flop operation in figure \ref{equiv}}
    \label{flopped}
\end{figure}
\squeezeup
The k\"ahler parameters of the pre-flopped geometry to the flopped geometry are related as 
\bea\label{eq:flop}
Q_1&=&Q^{\prime}_1Q_{\rho}^{\prime},\quad Q_2=Q^{\prime}_2Q_{\rho}^{\prime},\quad \tilde{Q}_1=\tilde{Q}^{\prime}_1Q_{\rho}^{\prime},\quad
 \tilde{Q}_2=\tilde{Q}^{\prime}_2Q_{\rho}^{\prime}\nonumber\\ Q_{\rho}&=&Q_{\rho}^{\prime-1},\quad
\eea
The crucial result that allows the use of flop transition to compute the open topological string amplitude is the flop invariance of topological string computations by refined vertex technique \cite{Taki:2008hb,Iqbal:2012xm,Sugimoto:2015nha}.  If we denote the CY3-folds corresponding to web diagrams in figure \ref{F0} and figure \ref{flopped}  as $X_1,X_2$, the flop invariance implies
\bea
&Z^{refined,open}_{X_2}(Q_1^{\prime},Q_2^{\prime},\tilde{Q}^{\prime}_1,\tilde{Q}^{\prime}_2,Q^{\prime}_\rho,q,t;x_i)&\nonumber\\=&Z^{refined,open}_{X_1}(Q^{\prime}_1Q_{\rho}^{\prime},Q^{\prime}_2Q_{\rho}^{\prime},\tilde{Q}_1,\tilde{Q}_2,Q_\rho,\tilde{Q}^{\prime}_1Q_{\rho}^{\prime},\tilde{Q}^{\prime}_2Q_{\rho}^{\prime},Q_{\rho}^{\prime-1},q,t;x_i)&\nonumber\\
\eea
We get the following refined amplitude for the flopped geometry in figure \ref{flopped} using the refined topological vertex formalism
\bea
Z_{open,X_2}^{ref}(Q_1,Q_2,\tilde{Q}_1,\tilde{Q}_2,Q_\rho,q,t;x)&=&\sum_{all\quad indices} (-Q_1)^{|\mu_1|}(-Q_2)^{|\mu_2|}(-\tilde{Q}_1)^{|\tilde{\mu}_1|}(-\tilde{Q}_2)^{|\tilde{\mu}_2|}(-Q_{\rho})^{|\rho|}\nonumber\\&\times&
C_{\nu^t\mu_2\tilde{\mu}_2}(t,q)C_{\rho\mu_2^t\tilde{\mu}^t_2}(q,t)C_{\rho^t\mu_1^t\tilde{\mu}_1}(t,q)C_{\emptyset\mu_1\tilde{\mu}^t_1}(q,t)s_{\nu^t}(x)
\eea
 Using the expression for the refined topological vertex  (\ref{eq:RTV2})
 and using   the  following skew Schur function identities repeatedly
\bea\label{schuridentity}
\sum_\lambda s_{\lambda/\alpha}(x)s_{\lambda/\beta}(y)&=&\prod_{i,j}(1-x_iy_j)^{-1}\sum s_{\beta/\eta}(x)s_{\alpha/\eta}(y)\nonumber\\
\sum_\lambda s_{\lambda^t/\alpha}(x)s_{\lambda/\beta}(y)&=&\prod_{i,j}(1+x_iy_j)\sum s_{\beta^t/\eta^t}(x)s_{\alpha^t/\eta}(y)
\eea
we get the expression 
\bea\label{eq:Zopen1}
&Z&_{open,X_2}^{ref}(Q_1,Q_2,\tilde{Q}_1,\tilde{Q}_2,Q_\rho,q,t;x)=\sum_{\tilde{\mu}_1,\tilde{\mu}_2}(-\tilde{Q}_1)^{|\tilde{\mu}_1|}(-\tilde{Q}_2)^{|\tilde{\mu}_2|} \tilde{Z}_{\tilde{\mu_1}}(t,q)\tilde{Z}_{\tilde{\mu^t_1}}(q,t)\tilde{Z}_{\tilde{\mu_2}}(t,q)\tilde{Z}_{\tilde{\mu^t_2}}(q,t)
\nonumber\\&\times&
\sqrt{t}^{||\tilde{\mu}^t_1||^2+||\tilde{\mu}^t_2||^2}\sqrt{q}^{||\tilde{\mu}_1||^2+||\tilde{\mu}_2||^2}
\prod(1-Q_1 q^{-\tilde{\mu_1}_i-\rho_j}t^{\tilde{-\mu^t_1}_j-\rho_i})\prod(1-Q_{\rho} q^{-\tilde{\mu_1}_j-\rho_i}t^{-\tilde{\mu^t_2}_i-\rho_j})\nonumber\\
&\times&\prod(1-Q_2 t^{-\tilde{\mu_2^t}_i-\rho_j}q^{-\tilde{\mu_2}_j-\rho_i})\prod(1-Q_1Q_{\rho} \sqrt{\frac{q}{t}}t^{\tilde{-\mu_2^t}_j-\rho_i}q^{-\tilde{\mu^t_1}_i-\rho_j})^{-1}\nonumber\\&\times&
\prod(1-Q_2Q_{\rho} \sqrt{\frac{t}{q}}t^{\tilde{-\mu_2^t}_j-\rho_i}q^{-\tilde{\mu_1}_i-\rho_j})^{-1}\prod(1-Q_1Q_2Q_{\rho}t^{\tilde{-\mu_2^t}_j-\rho_i}q^{-\tilde{\mu^t_1}_i-\rho_j})\nonumber\\
&\times&\prod(1+x_j  \sqrt{\frac{q}{t}} t^{-\rho_i}q^{-\tilde{\mu_2}_j})\prod(1+Q_2 \frac{q}{t}x_i t^{-\rho_j}q^{-\tilde{\mu_2}_j})^{-1}
\prod(1-Q_2Q_{\rho} \sqrt{\frac{q}{t}} x_it^{-\rho_j}q^{-\tilde{\mu_1}_j})^{-1}\nonumber\\&\times&\prod(1-Q_1Q_2Q_{\rho} \frac{q}{t}x_j t^{-\rho_i}q^{-\tilde{\mu^t_1}_i})
\eea
Since the refined topological vertex formalism gives both perturbative and non-perturbative parts from gauge theory view point, we have to normalise \footnote{Note that since for  $\tilde{Q}_1=\tilde{Q}_2=Q_b$ the exponent of $Q_b$ counts the instanton number, the gauge theory perturbative part is extracted by the limit $ Q_b\to 0$.} (\ref{eq:Zopen1}) to exclude the perturbative part. The normalised partition function turns out to be
\bea\label{eq:Znopen1}
&\hat{Z}&_{open,X_2}^{ref}(Q_1,Q_2,\tilde{Q}_1,\tilde{Q}_2,Q_\rho,q,t;x)=\sum_{\tilde{\mu}_1,\tilde{\mu}_2}(-\tilde{Q}_1)^{|\tilde{\mu}_1|}(-\tilde{Q}_2)^{|\tilde{\mu}_2|} \tilde{Z}_{\tilde{\mu_1}}(t,q)\tilde{Z}_{\tilde{\mu^t_1}}(q,t)\tilde{Z}_{\tilde{\mu_2}}(t,q)\tilde{Z}_{\tilde{\mu^t_2}}(q,t)\nonumber\\&\times&\sqrt{t}^{||\tilde{\mu}^t_1||^2+||\tilde{\mu}^t_2||^2}\sqrt{q}^{||\tilde{\mu}_1||^2+||\tilde{\mu}_2||^2}
\frac{\prod(1-Q_1 q^{-\tilde{\mu_1}_i-\rho_j}t^{\tilde{-\mu^t_1}_j-\rho_i})}{\prod(1-Q_1 q^{-\rho_j}t^{\tilde{-\rho_i}})}\frac{\prod(1-Q_{\rho} q^{-\tilde{\mu_1}_j-\rho_i}t^{\tilde{-\mu^t_2}_i-\rho_j})}{\prod(1-Q_{\rho} q^{-\rho_i}t^{\tilde{-\rho_j})}}\nonumber\\
&\times&\frac{\prod(1-Q_2 t^{\tilde{-\mu_2^t}_i-\rho_j}q^{-\tilde{\mu_2}_j-\rho_i})}{\prod(1-Q_2 t^{-\rho_j}q^{-\rho_i})}\frac{\prod(1-Q_1Q_{\rho} \sqrt{\frac{q}{t}}t^{-\rho_i}q^{-\rho_j})}{\prod(1-Q_1Q_{\rho} \sqrt{\frac{q}{t}}t^{\tilde{-\mu_2^t}_j-\rho_i}q^{-\tilde{\mu^t_1}_i-\rho_j})}\nonumber\\&\times&
\frac{\prod(1-Q_2Q_{\rho} \sqrt{\frac{t}{q}}t^{-\rho_i}q^{-\rho_j})}{\prod(1-Q_2Q_{\rho} \sqrt{\frac{t}{q}}t^{\tilde{-\mu_2^t}_j-\rho_i}q^{-\tilde{\mu_1}_i-\rho_j})}\frac{\prod(1-Q_1Q_2Q_{\rho}t^{\tilde{-\mu_2^t}_j-\rho_i}q^{-\tilde{\mu^t_1}_i-\rho_j})}{\prod(1-Q_1Q_2Q_{\rho}t^{-\rho_i}q^{-\rho_j})}\nonumber\\
&\times&\prod(1+x_j  \sqrt{\frac{q}{t}} t^{-\rho_i}q^{-\tilde{\mu_2}_j})\prod(1+Q_2 \frac{q}{t}x_i t^{-\rho_j}q^{-\tilde{\mu_2}_j})^{-1}
\prod(1-Q_2Q_{\rho} \sqrt{\frac{q}{t}} x_it^{-\rho_j}q^{-\tilde{\mu_1}_j})^{-1}\nonumber\\&\times&\prod(1-Q_1Q_2Q_{\rho} \frac{q}{t}x_j t^{-\rho_i}q^{-\tilde{\mu^t_1}_i})
\eea
A crucial step is to write the normalised partition function in terms of the K\"ahler parameters of the pre-flopped geometry  (\ref{F0}) using (\ref{eq:flop})
\bea\label{eq:Znopenpf}
&\hat{Z}&_{open,X_1}^{ref}(Q_1^{\prime},Q_2^{\prime},\tilde{Q}^{\prime}_1,\tilde{Q}^{\prime}_2,Q_\rho^{\prime},q,t;x)=\sum_{\tilde{\mu}_1,\tilde{\mu}_2}(-\tilde{Q}^{\prime}_1\tilde{Q}_{\rho}^{\prime})^{|\tilde{\mu}_1|}(-\tilde{Q}^{\prime}_2\tilde{Q}_{\rho}^{\prime})^{|\tilde{\mu}_2|} \tilde{Z}_{\tilde{\mu_1}}(t,q)\tilde{Z}_{\tilde{\mu^t_1}}(q,t)\tilde{Z}_{\tilde{\mu_2}}(t,q)\tilde{Z}_{\tilde{\mu^t_2}}(q,t)\nonumber\\&\times&\sqrt{t}^{||\tilde{\mu}^t_1||^2+||\tilde{\mu}^t_2||^2}\sqrt{q}^{||\tilde{\mu}_1||^2+||\tilde{\mu}_2||^2}
\frac{\prod(1-Q_1^{\prime}Q_{\rho}^{\prime} q^{-\tilde{\mu_1}_i-\rho_j}t^{\tilde{-\mu^t_1}_j-\rho_i})}{\prod(1-Q_1^{\prime}Q_{\rho}^{\prime} q^{-\rho_j}t^{\tilde{-\rho_i}})}\frac{\prod(1-(Q_{\rho}^{\prime})^{-1} q^{-\tilde{\mu_1}_j-\rho_i}t^{\tilde{-\mu^t_2}_i-\rho_j})}{\prod(1-(Q_{\rho}^{\prime})^{-1} q^{-\rho_i}t^{\tilde{-\rho_j})}}\nonumber\\
&\times&\frac{\prod(1-Q_2^{\prime}Q_{\rho}^{\prime} t^{\tilde{-\mu_2^t}_i-\rho_j}q^{-\tilde{\mu_2}_j-\rho_i})}{\prod(1-Q_2^{\prime}Q_{\rho}^{\prime} t^{-\rho_j}q^{-\rho_i})}\frac{\prod(1-Q_1^{\prime} \sqrt{\frac{q}{t}}t^{-\rho_i}q^{-\rho_j})}{\prod(1-Q_1^{\prime} \sqrt{\frac{q}{t}}t^{\tilde{-\mu_2^t}_j-\rho_i}q^{-\tilde{\mu^t_1}_i-\rho_j})}\nonumber\\&\times&
\frac{\prod(1-Q_2^{\prime} \sqrt{\frac{t}{q}}t^{-\rho_i}q^{-\rho_j})}{\prod(1-Q_2^{\prime} \sqrt{\frac{t}{q}}t^{\tilde{-\mu_2^t}_j-\rho_i}q^{-\tilde{\mu_1}_i-\rho_j})}\frac{\prod(1-Q_1^{\prime}Q_2^{\prime}Q_{\rho}^{\prime}t^{\tilde{-\mu_2^t}_j-\rho_i}q^{-\tilde{\mu^t_1}_i-\rho_j})}{\prod(1-Q_1^{\prime}Q_2^{\prime}Q_{\rho}^{\prime}t^{-\rho_i}q^{-\rho_j})}\nonumber\\
&\times&\prod(1+x_j  \sqrt{\frac{q}{t}} t^{-\rho_i}q^{-\tilde{\mu_2}_j})\prod(1+Q_2^{\prime}Q_{\rho}^{\prime} \frac{q}{t}x_i t^{-\rho_j}q^{-\tilde{\mu_2}_j})^{-1}\nonumber\\&\times&
\prod(1-Q_2^{\prime} \sqrt{\frac{q}{t}} x_it^{-\rho_j}q^{-\tilde{\mu_1}_j})^{-1}\prod(1-Q_1^{\prime}Q_2^{\prime}Q_{\rho}^{\prime} \frac{q}{t}x_j t^{-\rho_i}q^{-\tilde{\mu^t_1}_i})
\eea
It is important to note that the last expression has to be expanded in terms of $Q_{\rho}^{\prime}$ instead of $(Q_{\rho}^{\prime})^{-1}$ to prove its equivalence to (\ref{eq:Znopen1}).
Note that for particular values
\bea
Q_1^{\prime}Q_{\rho}^{\prime}=\sqrt{\frac{t}{q}},\nonumber\\
Q_2^{\prime}Q_{\rho}^{\prime}=\sqrt{\frac{t}{q}}
\eea
the expression (\ref{eq:Znopenpf}) reduces to
\bea\label{eq:correspondtoy1}
&Z_{open,X_1}^{ref}(Q_1^{\prime},Q_2^{\prime},\tilde{Q}^{\prime}_1,\tilde{Q}^{\prime}_2,Q_\rho^{\prime},q,t;x)&|_{Q_1^{\prime}Q_{\rho}^{\prime}=\sqrt{\frac{t}{q}},
Q_2^{\prime}Q_{\rho}^{\prime}=\sqrt{\frac{t}{q}}}=\nonumber\\&&\sum_{\tilde{\mu}_1,\tilde{\mu}_2}(-\tilde{Q}^{\prime}_1Q_{\rho}^{\prime})^{|\tilde{\mu}_1|}(-\tilde{Q}^{\prime}_2Q_{\rho}^{\prime})^{|\tilde{\mu}_2|}\sqrt{t}^{||\tilde{\mu}^t_1||^2+||\tilde{\mu}^t_2||^2}\sqrt{q}^{||\tilde{\mu}_1||^2+||\tilde{\mu}_2||^2}\nonumber\\
\eea
Moreover for the geometric engineering of pure $SU(2)$ with zero Chern-Simons coefficient for web diagram in figure (\ref{F0}), as in our case, we should impose the restrictions
\bea\label{eq:specialF0}
\tilde{Q}_1^{\prime}&=&\tilde{Q}_2^{\prime}=Q_b\nonumber\\
Q_1^{\prime}&=&Q_2^{\prime}=Q_f
\eea
 Restricting to the identification of parameters given in  (\ref{eq:specialF0}), corresponding to pure $SU(2)$ gauge theory, we get
\bea\label{eq:ZF0}
Z_{open}^{ref}(Q_b,Q_f,Q_m,q,t;x)&=&\sum_{\tilde{\mu}_1,\tilde{\mu}_2}(-Q_bQ_m)^{|\tilde{\mu}_1|}(-Q_bQ_m)^{|\tilde{\mu}_2|} \tilde{Z}_{\tilde{\mu_1}}(t,q)\tilde{Z}_{\tilde{\mu^t_1}}(q,t)\tilde{Z}_{\tilde{\mu_2}}(t,q)\tilde{Z}_{\tilde{\mu^t_2}}(q,t)\nonumber\\&\times&\sqrt{t}^{||\tilde{\mu}^t_1||^2+||\tilde{\mu}^t_2||^2}\sqrt{q}^{||\tilde{\mu}_1||^2+||\tilde{\mu}_2||^2}\nonumber\\&\times&\frac{\prod(1-Q_fQ_m q^{-\tilde{\mu_1}_i-\rho_j}t^{\tilde{-\mu^t_1}_j-\rho_i})}{\prod(1-Q_fQ_m q^{-\rho_j}t^{\tilde{-\rho_i}})}\frac{\prod(1-(Q_m)^{-1} q^{-\tilde{\mu_1}_j-\rho_i}t^{\tilde{-\mu^t_2}_i-\rho_j})}{\prod(1-(Q_m)^{-1} q^{-\rho_i}t^{\tilde{-\rho_j})}}\nonumber\\
&\times&\frac{\prod(1-Q_fQ_m t^{\tilde{-\mu_2^t}_i-\rho_j}q^{-\tilde{\mu_2}_j-\rho_i})}{\prod(1-Q_fQ_m t^{-\rho_j}q^{-\rho_i})}\frac{\prod(1-Q_f \sqrt{\frac{q}{t}}t^{-\rho_i}q^{-\rho_j})}{\prod(1-Q_f \sqrt{\frac{q}{t}}t^{\tilde{-\mu_2^t}_j-\rho_i}q^{-\tilde{\mu^t_1}_i-\rho_j})}\nonumber\\&\times&
\frac{\prod(1-Q_f \sqrt{\frac{t}{q}}t^{-\rho_i}q^{-\rho_j})}{\prod(1-Q_f \sqrt{\frac{t}{q}}t^{\tilde{-\mu_2^t}_j-\rho_i}q^{-\tilde{\mu_1}_i-\rho_j})}\frac{\prod(1-Q_fQ_fQ_mt^{\tilde{-\mu_2^t}_j-\rho_i}q^{-\tilde{\mu^t_1}_i-\rho_j})}{\prod(1-Q_fQ_fQ_mt^{-\rho_i}q^{-\rho_j})}\nonumber\\
&\times&\prod(1+x_j  \sqrt{\frac{q}{t}} t^{-\rho_i}q^{-\tilde{\mu_2}_j})\prod(1+Q_fQ_m \frac{q}{t}x_i t^{-\rho_j}q^{-\tilde{\mu_2}_j})^{-1}\nonumber\\&\times&
\prod(1-Q_f \sqrt{\frac{q}{t}} x_it^{-\rho_j}q^{-\tilde{\mu_1}_j})^{-1}\prod(1-Q_fQ_fQ_m \frac{q}{t}x_j t^{-\rho_i}q^{-\tilde{\mu^t_1}_i})
\eea
Similar to (\ref{eq:correspondtoy1}), choosing $Q_fQ_m=\sqrt{\frac{t}{q}}$ results in the simplifed expression
\bea\label{eq:Zrefy1}
Z_{open}^{ref}(Q_b,Q_f,Q_m,q,t;x)|_{Q_fQ_m=\sqrt{\frac{t}{q}}}&=&\sum_{\tilde{\mu}_1,\tilde{\mu}_2}(-Q_bQ_m)^{|\tilde{\mu}_1|+|\tilde{\mu}_2|}\sqrt{t}^{||\tilde{\mu}^t_1||^2+||\tilde{\mu}^t_2||^2}\sqrt{q}^{||\tilde{\mu}_1||^2+||\tilde{\mu}_2||^2}\nonumber\\
\eea
The topological string expression (\ref{eq:ZF0}) is to be compared with the  quiver moduli partition function (following (\ref{eq:Z5quiver}) and (\ref{eq:defect}) )  given as follows
\bea\label{eq:Z5pquiver}
&Z&^{quiver}_d(q_1,q_2,\rho_a,y,Q)=\sum_kQ^k\sum_{|\nu|=k}\nonumber\\&\times&
\frac{\prod_{(i,j)\in\nu^a}(1-yq_1^{(\nu^{b})^t_j}q_2^{j-\nu_i^a-1})\prod_{(i,j)\in\nu^b}(1-yq_1^{(\nu^{a})^t_j}q_2^{j-\nu_i^b-j})}{\prod_{(i,j)\in\nu^a}(1-q_1^{(\nu^{b})^t_j}q_2^{j-\nu_i^a-1})\prod_{(i,j)\in\nu^b}(1-q_1^{(\nu^{a})^t_j}q_2^{j-\nu_i^b-j})}
\nonumber\\&\times&
\sum_{\substack{(\underline{\mu},\underline{\nu})\\|\underline{\mu}|=|\underline{\nu}|+d}}
\Bigg[(\prod_{a=1}^2\prod_{i=1}^{c^a}\prod_{s=1}^{\mu^a_i-\nu^a_i}\rho_a q_1^{1-i}q_2^{-\nu^a_i-s+1})\prod_{a,b=1}^2\nonumber\\&\times&\bigg(\frac{\prod_{i=2}^{e^a+1}\prod_{j=1}^{c^b}\prod_{s=1}^{\mu^b_j-\nu^b_j}(1-y\rho_a\rho_b^{-1} q_1^{j-i}q_2^{\nu_j^b+s-\mu^a_i-1})}{\prod_{i=2}^{e^a+1}\prod_{j=1}^{c^b}\prod_{s=1}^{\mu^b_j-\nu^b_j}(1-\rho_a\rho_b^{-1} q_1^{j-i}q_2^{\nu_j^b+s-\mu^a_i-1})}
\frac{\prod_{j=1}^{c^b}\prod_{s=1}^{\mu^b_j-\nu^b_j}(1-\rho_a\rho_b^{-1}y q_1^{j-1}q_2^{-\nu^b_j+s-\mu^a_1-1})}{\prod_{j=1}^{c^b}\prod_{s=1}^{\mu^b_j-\nu^b_j}(1- \rho_a\rho_b^{-1}q_1^{j-1}q_2^{-\nu^b_j+s-\mu^a_1-1})}\nonumber\\ &\times&
\frac{\prod_{i=2}^{e^a+1}\prod_{j=1}^{c^b}\prod_{r=0}(1-y \rho_a\rho_b^{-1}q_1^{j-i}q_2^{\nu_j^b-\nu_{i-1}^a-r-1})(1-y \rho_a\rho_b^{-1}q_1^{j-i}q_2^{\mu_j^b-\nu_{i-1}^a+r})}{\prod_{i=2}^{e^a+1}\prod_{j=1}^{c^b}\prod_{r=0}(1- \rho_a\rho_b^{-1}q_1^{j-i}q_2^{\nu_j^b-\nu_{i-1}^a-r-1})(1- \rho_a\rho_b^{-1}q_1^{j-i}q_2^{\mu_j^b-\nu_{i-1}^a+r})}\bigg)
\Bigg]\nonumber\\
\eea
\bea\label{eq:Zquiverry1}
Z^{quiver}_d(q_1,q_2,\rho_a,y,Q)|_{y=1}&=&\sum_kQ^k\sum_{|\underline{\nu}|=k}\sum_{\substack{(\underline{\mu},\underline{\nu})\\|\underline{\mu}|=|\underline{\nu}|+d}}
\Bigg[(\prod_{a=1}^2\prod_{i=1}^{c^a}\prod_{s=1}^{\mu^a_i-\nu^a_i}\rho_a q_1^{1-i}q_2^{-\nu^a_i-s+1})
\Bigg]\nonumber\\
\eea
\section{Relations between the parameters in the conjecture}\label{DTKT}
In this section we describe  consistency conditions that  lead to the complete identification of parameters  from the two sides of the conjectures (\ref{eq:5dconj},\ref{eq:chiyr2}).  We will also formulate the conjecture in a more general form that contains information about the holonomy observables, denoted by $\bf{x}$, parametrising the lagrangian branes. For  the unrefined case in the absence of  lagrangian branes on the external toric legs see \cite{Chuang:2013wpa,Chuang:2012dv}.
Note the following facts: 
\begin{itemize}
\item In the decompactification limit $-log(Q_1)\to \infty$ the 4d version of   (\ref{eq:5dconj})  imposes the following identifications \cite{Bruzzo:2010fk}
\end{itemize}
\bea
t=q_1,\quad q=q_2^{-1},\quad Q_2=T\sqrt{q_1q_2}
\eea
\begin{itemize}
\item By considering the special case $\nu=\emptyset,y=1$ in (\ref{eq:5dconj})  one must choose $Q_1=\sqrt{\frac{t}{q}}y$  for the identity to hold. Since from $2d$ sigma model point of view $y$ is the mass parameter, this identification of parameters does not change for  $\nu\ne\emptyset$. 
\end{itemize}
\begin{itemize}
\item As we consider cases $r\ge 2$ the characters $\rho_a,$ are related to the K\"ahler parameters $-log(Q_{f_i}),-log(Q_b)$ of the fiber  and base directions .
\end{itemize}
Given these constraints , the generalisation of (\ref{eq:5dconj}) to the $rank=2$ case  will then be given by
\bea\label{eq:chiyr2}
Z^{quiver}_d(q_1,q_2,\rho_a,y,Q)=_{parameter-identification}\tilde{Z}^{ref}_{open,d}(Q_f,Q_b,Q_m,q,t)
\eea
Note that after parameter identification $q=q_2^{-1}, t=q_1,Q_f=\rho_1^{-1}\rho_2$,  the decompactification limit $-log(Q_m)\to \infty$ leads to  the result (5.3) proved in \cite{Bruzzo:2010fk}
\bea
Z^{quiver}_d(q_1,q_2,\rho_1,\rho_2,Q)=\tilde{Z}_{open,d}(q_1,q_2,\rho_1^{-1}\rho_2,Q)
\eea
\subsection{Unrefined topological strings}
A subclass of the quiver moduli spaces discussed in section (\ref{section2}) and its corresponding partition function was discussed in detail by Diconescu et al. \cite{Chuang:2012dv,Chuang:2013wpa}. This subclass of moduli spaces does not contain the moduli corresponding to the D4-branes, the open strings between D4-D6 branes and D4-D2 branes.  This moduli space modulo certain equivalence relations was shown to be isomorphic to the nested Hilbert scheme of points on $\mathbb{C}^2$.  This scheme, denoted by $\mathcal{N}(\gamma)$ depends on an ordered sequence $\gamma=\{m_{a_i} \}_{0\le i\le k}$ of positive integers and parametrize a sequence of ideals sheaves $0\subset\mathcal{I}_k\subset\mathcal{I}_{k-1\subset...\subset\mathcal{I}_0}$ of zero dimensional  subschemes $Z_i\subset\mathbb{C}^2$ and the corresponding topological data $\chi(\mathcal{O}_{Z_i})=\sum_{j=0}^i\gamma_j$ for $0\le i\le l$.\\
The character valued partition function is given by the equivariant $\chi_y$ genus of a bundle $\mathcal{V}$ on  $\mathcal{N}(\gamma)$. The bundle is identical to the bundle $\mathcal{L}_{(g,p)}$  described earlier and given as
\bea
\mathcal{V}(\gamma)\equiv \eta^{*}(\mathcal{V}_{g,p}) \simeq\eta^{*}\big(T^{*}\mathcal{H}^{r\oplus g}\otimes det (\mathbb{V})^{p}\big)
\eea
The existence of the morphism $\eta:\mathcal{N}(\gamma)\to\mathcal{H}^r$ to the Hilbert scheme of $r$ points on $\mathbb{C}^2$ makes it possible to apply the equivariant localization. The fixed points are given by the monomial ideals that are in one-to-one correspondence with the partitions of $r$.\\
More interesting is the appearance of the modified Kostka-Macdonald coefficients. It can be explained by the existence of a map from the nested Hilbert scheme\\ $\rho:\mathcal{N}(1,1,...,1)\footnote{the number of $1$s in $(1,1,...,1)$ is equal to $r$.}\to\tilde{\mathcal{H}}^r$ to isospectral Hilbert scheme discussed in \cite{haiman2000hilbert}. The web of maps is shown below in the figure (\ref{isospectral}) as a commutative diagram. With this commutative diagram in mind, it was shown that there exists two very important pushforward maps
\bea\label{eq:pushforward}
\rho^{\gamma}_{red *}\mathcal{O}_{\mathcal{N}_{(\gamma)}}&=&\mathcal{O}_{\tilde{\mathcal{H}_{red}^{\gamma}}}\nonumber\\
\pi_{red*}^{\gamma}\mathcal{O}_{\tilde{\mathcal{H}}_{red}}&\simeq& (\pi_{red*}
\mathcal{O}_{\tilde{\mathcal{H}_{red}}})^{S_{\gamma}}=\mathcal{P}^{S_{\gamma}}
\eea
Next we summarise a sequence of arguments that fixes the notation of this section and which lead to the expansion of the topological string partition function in terms of the modified Macdonald polynomials. \\
a)since $\mathcal{N}(1,1,...,1)$ is reduced, this implies the existence of the morphism \\ $\rho_{red}:\mathcal{N}(1,1,...,1)\to \tilde{\mathcal{H}}^r_{red}$, \\
b)$\rho_{red*}\mathcal{O}_{\mathcal{N}(1,1...,1)}\equiv \mathcal{O}_{\tilde{\mathcal{H}}^r_{red}}$,\\
c)the pushforward  map $\pi_{red*}\mathcal{O}_{\tilde{\mathcal{H}}_{red}}$ is a vector bundle $\mathcal{P}$ on the Hilbert scheme and is isomorphic to the pushforward map $\eta_{*}\mathcal{O}_{\mathcal{N}(1,1,...,1)}$,\\
d)consider the stabiliser $S_{\gamma}\subset S_r$ of the partition $\gamma$ with $S_r$ the symmetric group of order $r$. This shows that $\tilde{\mathcal{H}}^r$ furnishes a representation of $S_{\gamma}$ by the restriction map, $S_{\gamma}\times \tilde{\mathcal{H}}^r\to \tilde{\mathcal{H}}^r$,\\
e)$\tilde{\mathcal{H}}^{\gamma}$ denotes the quotient of $\tilde{\mathcal{H}}^r$ by $S_{\gamma}$,\\
f)there exists a morphism $\rho^{\gamma}:\mathcal{N}(\gamma)\to\tilde{\mathcal{H}}^{\gamma}$ which is also true for the corresponding reduced schemes $\rho^{\gamma}_{red}:\mathcal{N}(\gamma)\to\tilde{\mathcal{H}}^{\gamma}_{red}$.\\
As a consequence of the equations (\ref{eq:pushforward}) in the  $\bf{T}$-equivariant framework
we have 
\bea\label{eq:chiyloca}
\chi_{y}^T(\mathcal{N}(\gamma),\eta^{\gamma *}\mathcal{V}_{g,p})&=&\chi_{y}^T(\mathcal{H}^r,(\mathcal{P}^{S_{\gamma}}\times_{\mathcal{H}^r}\mathcal{V}_{g,p}))\nonumber\\&=&\sum_{\mu}\Omega_{\mu}^{g,p}(q_1,q_2,y)ch_T(\mathcal{P}_{\mu}^{\gamma})=\sum_{\mu}\Omega_{\mu}^{g,p}(q_1,q_2,y)\sum_{\lambda}K_{\lambda,\tilde{\gamma}}\tilde{K}_{\lambda,\mu}(q_1,q_2)\nonumber\\
\eea
where $\tilde{\gamma}$ denotes an unordered partition of $r$ determined by the sequence $\gamma$, $K_{\lambda,\tilde{\gamma}}$ are the Kostka numbers and $\tilde{K}_{\lambda,\mu}(q_1,q_2)$ are the modified Kostka-Macdonald coefficients and in the second last equality equivariant localization was used. The character valued (K-theoretic) partition function is then given by the generating function of  the $\chi_y$ genus
\bea
Z_{K}^r(q_1,q_2,y;\mbox{\bf{x}})=\sum_{\substack {m_0+m_1+...+m_{r-1}=r \\ m_i\in\mathbb{Z}_{\ge 0}}}\chi_{y}^T(\mathcal{N}(\gamma),\eta^{\gamma *}\mathcal{V}_{g,p})\prod_{i=0}^{r-1}x_i^{m_a}
\eea
where the factor $\prod_{i=0}^{r-1}x_i^{m_a}$ denotes the expansion of the partition function in terms of the symmetric functions of the holonomy observables.\\
Using the equation (\ref{eq:chiyloca}) one gets
\bea
Z_{K}^r(q_1,q_2,y;\mbox{\bf{x}})=\sum_{\mu}\Omega_{\mu}^{g,p}(q_1,q_2,y)\tilde{H}_{\mu}(q_2,q_1;\mbox{\bf{x}})
\eea
For the moduli space of this section .i.e. $\mathcal{H}^r(\mathbb{C}^2)$ the equivariant localization yields
\bea
\Omega_{\mu}^{g,p}(q_1,q_2,y)=\prod_{\mu}(q_1^{l(\square)}q_2^{a(\square)})^{g-1-p}\frac{(1-yq_1^{-l(\square)}q_2^{a(\square)+1})^g(1-yq_1^{l(\square)+1}q_2^{-a(\square)})^g}{(1-q_1^{-l(\square)}q_2^{a(\square)+1})(1-q_1^{l(\square)+1}q_2^{-a(\square)})}
\eea
with $a(\square)$ and $ l(\square)$ defined as the arm length and the leg length of a box in the Young diagram corresponding to $\mu$. Therefore
\bea
Z_{K}^r(q_1,q_2,y;\mbox{\bf{x}})=\sum_{\mu}\prod_{\mu}(q_1^{l(\square)}q_2^{a(\square)})^{g-1-p}\frac{(1-yq_1^{-l(\square)}q_2^{a(\square)+1})^g(1-yq_1^{l(\square)+1}q_2^{-a(\square)})^g}{(1-q_1^{-l(\square)}q_2^{a(\square)+1})(1-q_1^{l(\square)+1}q_2^{-a(\square)})}\tilde{H}_{\mu}(q_2,q_1;\mbox{\bf{x}})\nonumber\\
\eea
As shown in \cite{Chuang:2010ii,Chuang:2012dv} a particular change of variables \footnote{Unfortunately this change of variables does not have a conceptual derivation. It was worked out by the requirement that the following conjecture should hold: \nonumber\\
 DT partition function of the CY3-fold X=Instanton partition function of 5d SUSY gauge theory geometrically engineered by X.} motivated by the geometric engineering conjecture between the 5d supersymmetric gauge theory with eight super charges and the Donaldson thomas theory of CY3-fold X, relates the K-theoretic partition function to the unrefined open string invariants as
\bea\label{eq:DTBPs}
Z^{top,open}_X(q,\mbox{\bf{x}},y)&=^{?}&1+\sum_{r\ge 1}Z_K^r(qy^{-1},q^{-1}y^{-1},(-1)^{g-1-p}y^{-g}\mbox{\bf{x}})\nonumber\\
&=&1+\sum_{\mu\ne\emptyset}\prod_{\square\in\mu}(q^{l(\square)-a(\square)}y^{-l(\square)-a(\square)})^p(qy^{-1})^{2l(\square)+2a(\square)} (-1)^{p|\mu|}\nonumber\\
&\times&\frac{(1-y^{l(\square)-a(\square)}q^{-a(\square)-l(\square)-1})^{2g}}{(1-y^{l(\square)-a(\square)-1}q^{-a(\square)-l(\square)-1})(1-y^{l(\square)-a(\square)+1}q^{-a(\square)-l(\square)-1})}\nonumber\\&\times&\tilde{H}_{\mu}(q^{-1}y^{-1},qy^{-1},\bf{x})\nonumber\\
\eea
The quantity $Z^{top,open}_X(q,\mbox{\bf{x}},y)$ can be independently computed using the topological vertex formalism. It turns out \cite{Chuang:2012dv}  e.g. for $\mbox{\bf{x}}=\{Q,0,0.,,,\}$, to be equal to the right hand side of (\ref{eq:DTBPs}) for all allowed values of $g$ and $p$
\bea
&1+\sum_{\mu\ne\emptyset}\prod_{\square\in\mu}(q^{l(\square)-a(\square)}y^{-l(\square)-a(\square)})^p(qy^{-1})^{2l(\square)+2a(\square)} (-1)^{p|\mu|}\nonumber\\
&\times\frac{(1-y^{l(\square)-a(\square)}q^{-a(\square)-l(\square)-1})^{2g}}{(1-y^{l(\square)-a(\square)-1}q^{-a(\square)-l(\square)-1})(1-y^{l(\square)-a(\square)+1}q^{-a(\square)-l(\square)-1})}Q^{|\mu|}\nonumber\\
&=\sum_{\mu}(-1)^{p|\mu|}q^{-(g-1-p)\kappa(\mu)}(\prod_{\square\in\mu}(q^{\frac{a(\square)+l(\square)}{2}}-q^{-\frac{a(\square)+l(\square)}{2}})^{2g-2})Q^{|\mu|}
\eea
where $\kappa(\mu)=\sum_{\square\in\mu}(i(\square)-j(\square))$.\\
For $y=1$ it was indicated that the following identity is crucial for proving  the last conjectural equality
\bea
\sum_{\square\in\mu}(l(\square)-a(\square))=\sum(j(\square)-i(\square))
\eea
\begin{figure}
\begin{tikzpicture}
\begin{tikzcd}
\mathcal{N}(1,1,...,1)\arrow{r} \arrow[swap]{d}{\rho} &(\mathbb{C}^2)^r\arrow{d}{Identity}  \\
\tilde{\mathcal{H}}^r \arrow{r} \arrow[swap]{d}{\pi} & (\mathbb{C}^2)^r \arrow{d} \\%
\mathcal{H}^r  \arrow{r}& S^r(\mathbb{C}^2)
\draw[ thick, ->] (-3.1,2) arc (90:290:1);
\node [left] at (-4.2,0.7) {$\eta$};
\end{tikzcd}
\end{tikzpicture}
\caption{}
    \label{isospectral}
\end{figure}
\subsection{Refined topological strings}
The main purpose of this work is to give the generalisations of these conjectural identities for the refined topological string case.
\bea
&Z&^{5d,instanton}(q_1,q_2,\rho_a,y,Q,\bf{x})=\sum_kQ^k\sum_{|\underline{\nu}|=k}
\prod_{a,b=1}^r\nonumber\\&\times&
\frac{\prod_{(i,j)\in\nu^a}(1-y\rho_a\rho_b^{-1}q_1^{(\nu^{b})^t_j}q_2^{j-\nu_i^a-1})\prod_{(i,j)\in\nu^b}(1-y\rho_a\rho_b^{-1}q_1^{(\nu^{a})^t_j}q_2^{j-\nu_i^b-j})}{\prod_{(i,j)\in\nu^a}(1-\rho_a\rho_b^{-1}q_1^{(\nu^{b})^t_j}q_2^{j-\nu_i^a-1})\prod_{(i,j)\in\nu^b}(1-\rho_a\rho_b^{-1}q_1^{(\nu^{a})^t_j}q_2^{j-\nu_i^b-j})}\nonumber\\&\times&
\sum_{\substack{(\underline{\mu},\underline{\nu})\\|\underline{\mu}|=|\underline{\nu}|+d}}
\Bigg[(\prod_{a=1}^r\prod_{i=1}^{c^a}\prod_{s=1}^{\mu^a_i-\nu^a_i}\rho_a q_1^{1-i}q_2^{-\nu^a_i-s+1})\nonumber\\&\times&\frac{\prod_{a,b=1}^r\prod_{i=2}^{e^a+1}\prod_{j=1}^{c^b}\prod_{s=1}^{\mu_j^b-\nu_j^b}(1-y\rho_a\rho_b^{-1} q_1^{j-i}q_2^{\nu^b_j+s-\mu^a_i-1})}{\prod_{a,b=1}^r\prod_{i=2}^{e^a+1}\prod_{j=1}^{c^b}\prod_{s=1}^{\mu_j^b-\nu_j^b}(1-\rho_a\rho_b^{-1} q_1^{j-i}q_2^{\nu^b_j+s-\mu^a_i-1})}\nonumber\\ &\times&
\frac{\prod_{a,b=1}^r\prod_{i=2}^{e^a+1}\prod_{j=1}^{c^b}\prod_{s=1}^{\mu^b_j-\nu^b_j}(1-\rho_a\rho_b^{-1} q_1^{j-i}q_2^{\nu^b_j+s-\nu^a_{i-1}-1})}{\prod_{a,b=1}^r\prod_{i=2}^{e^a+1}\prod_{j=1}^{c^b}\prod_{s=1}^{\mu^b_j-\nu^b_j}(1-y\rho_a\rho_b^{-1} q_1^{j-i}q_2^{\nu^b_j+s-\nu^a_{i-1}-1})}\nonumber\\&\times&
\frac{\prod_{a,b=1}^r\prod_{j=1}^{c^b}\prod_{s=1}^{\mu^b_j-\nu^b_j}(1-y \rho_a\rho_b^{-1}q_1^{j-1}q_2^{-\nu^b_j+s-\mu^a_1-1})}{\prod_{a,b=1}^r\prod_{j=1}^{c^b}\prod_{s=1}^{\mu^b_j-\nu^b_j}(1- \rho_a\rho_b^{-1}q_1^{j-1}q_2^{-\nu^b_j+s-\mu^a_1-1})}\Bigg]\tilde{H}_{\underline{\mu}}(q_1,q_2,\bf{x})\nonumber\\
\eea
To substantiate the conjecture (\ref{eq:DTBPs}) for the refined topological strings in the rank $r=1,2$ cases we have to find a map between the chemical potentials from the two sides.
A natural generalisation for spacetime equivariant parameters is 
\bea
q_1&=&ty^{-1},\quad q_2=q^{-1}y^{-1},\quad Q_2=T\sqrt{\frac{t}{q}}y^{-1}\nonumber\\
Q_1&=&\sqrt{\frac{t}{q}} y,\quad Q_{f_i}=\rho_i\rho_{i+1}^{-1}
\eea
Making this change of variables conjecturally identifies the $5d$ Nekrasov partition function with the refined open topological string partition function
\bea
&Z^{ref,open}(q,t,Q_1,Q_2,Q_{f_i};\mbox{\bf{x}})=Z^{5d,instanton}(q_1,q_2,\rho_a,y,T;\mbox{\bf{x}})\Bigg\rvert_{\{q_1=ty^{-1},\quad q_2=q^{-1}y^{-1},\quad Q_2=T\sqrt{\frac{t}{q}}y^{-1}
Q_1=\sqrt{\frac{t}{q}} y,\quad Q_{f_i}=\rho_i\rho_{i+1}^{-1}
\}}
\eea
\subsection{Lagrangian branes along the un-preferred vs preferred direction:Schur  polynomials vs Macdonald polynomials  }
The Schur polynomials and the Macdonald polynomials are two of the symmetric functions bases in which we can expand
he refined open topological string partition functions. The  choice of the Schur polynomials corresponds \cite{Kozcaz:2018ndf} to the lagrangian branes present on the unpreferred leg of the toric diagram, whereas the Macdonald polynomials basis corresponds to the lagrangian brane on the preferred leg of the toric diagram. Interestingly the modified Macdonald polynomials can be expanded in terms of both the Schur functions and the Macdonald polynomials as follows \cite{Chuang:2013wpa,Haglund_2004,Haglund_2004}
\bea
\tilde{H}_{\mu}(\mbox{\bf{x}};q,t)&=&\sum_{\lambda}\tilde{K}_{\lambda\mu}(q,t)s_{\lambda}(\mbox{\bf{x}})\nonumber\\
\tilde{H}_{\mu}(\mbox{\bf{x}};q,t)&=& t^{-\sum_i(\mu_i(i-1))}J_{\mu}\big[\frac{\mbox{\bf{x}}}{1-t^{-1}};q,t^{-1} \big]\nonumber\\
&=& t^{-\sum_i(\mu_i(i-1))}\prod_{s\in (\lambda)}(1-q^{a(s)}t^{l(s)+1})P_{\lambda}(\mbox{\bf{x}};q,t)
\eea
where $\tilde{K}_{\lambda\mu}(q,t)$ are the modified Kostka coefficients and have interesting combinatorial properties, $s$ specifies a box in the Young diagram, $a(s)$ and $l(s)$ denote the arm length  and the leg length of the square $s$ repsectively, $J_{\lambda}(\mbox{\bf{x}};q,t)$ is defined as the integral form of the Macdonald polynomials 
\bea
J_{\lambda}(\mbox{\bf{x}};q,t):=\prod_{s\in D(\lambda)}(1-q^{a(s)})t^{l(s)+1})P_{\lambda}(\mbox{\bf{x}};q,t) 
\eea
and 
\bea
J_{\mu}\big[\frac{\mbox{\bf{x}}}{1-t^{-1}};q,t^{-1} \big]:=J_{\mu}\big[x_1,x_2,...,t^{-1}x_1,t^{-1}x_2,...,t^{-2}x_1,t^{-2}x_2,...;q,t^{-1} \big]. 
\eea
A crucial point is that  the  choice of the preferred or un preferred direction for the placement of the lagrangian branes corresponds to the expansion of the modified Macdonald polynomials in the Macdonald polynomials basis or the Schur functions basis.
\section{Elliptic genus: a speculation for the refined open string invariants of special lagrangian branes for fully compactified web}\label{elliptge2}
In the same vein as the purported equality of the Donaldson-Thomas partition function of the CY3-fold and the K-theory partition function of the framed quiver moduli space given in the last section, we give an expression for the generating function of the elliptic genus of the same framed moduli space of section \ref{section2} and propose that 
\bea
&The\quad Donaldson-Thomas\quad partition \quad functions \quad of\quad the\quad CY3-folds (fig.\ref{fullcompAnfibP1})\nonumber\\ &= \nonumber\\ &The \quad generating\quad function\quad of\quad the \quad elliptic\quad genus\quad given\quad in\quad (\ref{eq:GEG1},\ref{eq:GEG2})\nonumber\\
\eea
In the presence of vector bundles on quiver moduli space $\mathcal{N} (r, k + d, d)$, the natural generalisation of the $\chi_y$-genus is the so-called elliptic genus. The elliptic genus contains topological information about the vector bundles and can be arranged as a generating series of cohomology groups  of the vector bundles. To define it, consider a vector bundle $V$ on $X$ and define the formal product 
\bea
\mathcal{E}(V)=\otimes_{n\ge 1} (\Lambda_{-yq^n}V^{\vee}\otimes\Lambda_{-y^{-1}q^n}\otimes S_{q^n}(V\oplus V^{\vee}))
\eea
It is interesting to note that this formal product is the vector bundle analogue of the Jacobi triple product formula\cite{Gritsenko:1999nm,Haghighat:2013gba}.
 The elliptic genus $\chi_{elliptic}(X,V;y,q)$  is defined by its relation to the chi-y genus as
\bea
\chi_{elliptic}(X,V;y,q):=y^{-rkT_X/2}\chi_{-y}(X,\mathcal{E}(T_X)\otimes V)
\eea
Then using  Riemann-Roch theorem the elliptic genus $\chi_{elliptic}(X,V;y,q)$ can be expressed by
\bea
\chi_{elliptic}(X,V;y,q)&=&\int_{[X]} y^{-rkT_X/2}\mbox{ch}(\Lambda_{-y}T_X^{\vee})\mbox{ch}(\mathcal{E}(T_X))\mbox{td}(T_X)\mbox{ch}(V)\nonumber\\
&=&\int_X((\sum_{l=1}^de^{v_k}))\prod_{i=1}^nx_i\frac{\theta(\frac{x_i}{2\pi i}-z,\tau)}{\theta(\frac{x_i}{2\pi i},\tau)}\prod_{i=1}^m\frac{\theta(\frac{u_i}{2\pi i},\tau)}{\theta(\frac{u_i}{2\pi i}-z,\tau)}
\eea
where $q=e^{2\pi i\tau},y=e^{2\pi i z}$ and $\theta(z,\tau)=q^{1/8}\frac{y^{1/2}-y^{-1/2}}{i}\prod_{l=1}(1-q^l)(1-q^ly)(1-q^ly^{-1})$.  For  the moduli space $\mathcal{N} (r, n + d, d)$ one has to generalise the above definitions to include virtual schemes allowing an equivariant torus action and use equivariant localisation. We follow the fixed point formulas given in these references  to write down the final expressions for $\chi_y$ genus and elliptic genus of $\mathcal{N} (r, n + d, d)$. Note that $T_X$ for $X=\mathcal{N} (r, n + d, d)$ is given in (\ref{fixedpointset}). For a quick review of the fixed point formulae see appendix (\ref{virtuallocalization}).   \\
 The computation of  the refined open topological string amplitude corresponding to the Euler characteristic and $\chi_y$ genus  was relatively simple in the sense that the lagrangian brane was put on the external leg of the toric diagram. So it was a topological amplitude in the presence of the external branes. In the case of a totally compactified web diagram, all the legs are essentially internal.   The instanton partition function of the six dimensional theory with surface defect  can be interpreted as the generating function of  the elliptic genus
 \bea
Z^{6d,quiver}_{r,d}=\sum_k Q_{\rho}^k \chi_{ell}(\mathcal{N} (r, k + d, d),q_1,q_2,y,\rho_a,Q_{\sigma})
\eea  
\\
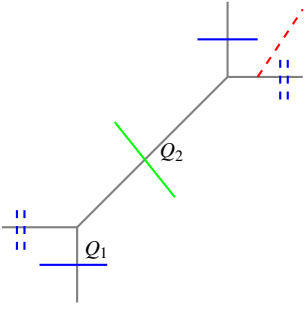
\begin{figure}
\begin{tikzpicture}
\draw[gray, thick] (2,4) -- (3,4);
\draw[red, thick,dashed] (2.4,4)-- (3,4.9);
\draw[blue, thick,dashed] (2.7,3.7)-- (2.7,4.3);
\draw[blue, thick,dashed] (2.8,3.7)-- (2.8,4.3);
\draw[gray, thick] (2,4) -- (2,5);
\draw[blue, thick] (1.6,4.5) -- (2.4,4.5);
\draw[gray, thick] (0,2) -- (2,4);
\draw[green, thick]  (0.5,3.4) -- (1.3,2.4);
\node [right] at (1,3) {$Q_2$};
\draw[gray, thick] (0,2) -- (0,1);
\draw[blue, thick] (-0.5,1.5) -- (0.4,1.5);
\draw[gray, thick] (0,2) -- (-1,2);
\draw[blue, thick,dashed] (-0.7,1.7)-- (-0.7,2.3);
\draw[blue, thick,dashed] (-0.8,1.7)-- (-0.8,2.3);
\node [right] at (0.0,1.7){$Q_1$};
\end{tikzpicture}
\caption{totally compactified toric web of the the total space of the bundle $\mathcal{O}(-1)\oplus \mathcal{O}(-1) \to \mathbb{P}^1$}
    \label{fullcompconifold}
\end{figure}
 \\ $Z^{6d,quiver}_{r,d}$ will provide the prediction for $\tilde{Z}^{ref}_{open,d}$ for general rank $r$ and a fully compactified web.
In figure \ref{fullcompconifold} we give the totally compactified toric web of the the total space of the bundle $\mathcal{O}(-1)\oplus \mathcal{O}(-1) \to \mathbb{P}^1$.
\bea\label{eq:GEG1}
Z^{6d,quiver}_{1,d}=
&\sum_k Q_{\rho}^k \chi_{ell}(\mathcal{N} (1, k + d, d),q_1,q_2,y,\rho_1,Q_{\sigma})=
\sum_{|\underline{\nu}|=k}Q_{\rho}^k\times\nonumber\\&
\prod_{(i,j)\in\nu}\frac{(1-yq_1^{(\nu^{})^t_j}q_2^{j-\nu_i-1})}{(1-q_1^{(\nu^{})^t_j}q_2^{j-\nu_i-1})}\prod_{k=1}\frac{(1-y Q_{\sigma}^kq_1^{(\nu^{1})^t_j}q_2^{j-\nu_i-1})(1-y^{-1} Q_{\sigma}^kq_1^{-(\nu^{})^t_j}q_2^{-j+\nu_i+1})}{(1-Q_{\sigma}^kq_1^{(\nu^{1})^t_j}q_2^{j-\nu_i-1})(1-Q_{\sigma}^kq_1^{-(\nu^{1})^t_j}q_2^{-j+\nu_i+1})}\nonumber\\&
\prod_{(i,j)\in\nu}\frac{(1-yq_1^{(\nu^{})^t_j}q_2^{j-\nu_i-j})}{(1-q_1^{(\nu^{})^t_j}q_2^{j-\nu_i-j})}\prod_{k=1}\frac{(1-y Q_{\sigma}^kq_1^{(\nu^{})^t_j}q_2^{j-\nu_i-j})(1-y^{-1} Q_{\sigma}^kq_1^{-(\nu^{1})^t_j}q_2^{-j+\nu_i+j})}{(1-Q_{\sigma}^kq_1^{(\nu^{})^t_j}q_2^{j-\nu_i-j})(1-Q_{\sigma}^kq_1^{-(\nu^{})^t_j}q_2^{-j+\nu_i+j})}
\nonumber\\&\sum_{\substack{(\mu,\nu)\\|\mu|=|\nu|+d}}
\Bigg[\bigg(\prod_{i=1}^{c}\prod_{s=1}^{\mu_i-\nu_i}\rho_1 q_1^{1-i}q_2^{-\nu_i-s+1}\bigg)\nonumber\\&\times
\nonumber\\&\bigg(\prod_{i=2}^{e+1}\prod_{j=1}^{c}\prod_{s=1}^{\mu_j-\nu_j}\nonumber\\&\frac{(1-yq_1^{j-i}q_2^{\nu_j+s-\mu_i-1})}{(1- q_1^{j-i}q_2^{\nu_j+s-\mu_i-1})}\prod_{k=1}\frac{(1-y Q_{\sigma}^kq_1^{j-i}q_2^{\nu_j+s-\mu_i-1})(1-y^{-1} Q_{\sigma}^k q_1^{-j+i}q_2^{-\nu_j-s+\mu_i+1})}{(1-Q_{\sigma}^k q_1^{j-i}q_2^{\nu_j+s-\mu_i-1})(1-Q_{\sigma}^k q_1^{-j+i}q_2^{-\nu_j-s+\mu_i+1})}\bigg)\nonumber\\ &
\bigg(\frac{\prod_{i=2}^{e+1}\prod_{j=1}^{c}\prod_{p=0}(1-y q_1^{j-i}q_2^{\nu_j-\nu_{i-1}-p-1})(1-yq_1^{j-i}q_2^{\mu_j-\nu_{i-1}+p})}{\prod_{i=2}^{e+1}\prod_{j=1}^{c}\prod_{p=0}(1- q_1^{j-i}q_2^{\nu_j-\nu_{i-1}-p-1})(1- q_1^{j-i}q_2^{\mu_j-\nu_{i-1}+p})}\nonumber\\&
\frac{\prod_{i=2}^{e+1}\prod_{j=1}^{c}\prod_{p=0}(1-yQ_{\sigma}^k q_1^{j-i}q_2^{\nu_j-\nu_{i-1}-p-1})(1-y Q_{\sigma}^k q_1^{j-i}q_2^{\mu_j-\nu_{i-1}+p})}{\prod_{i=2}^{e+1}\prod_{j=1}^{c}\prod_{p=0}(1-Q_{\sigma}^k q_1^{j-i}q_2^{\nu_j-\nu_{i-1}-p-1})(1- Q_{\sigma}^kq_1^{j-i}q_2^{\mu_j-\nu_{i-1}+p})}\nonumber\\&
\frac{\prod_{i=2}^{e+1}\prod_{j=1}^{c}\prod_{p=0}(1-y^{-1}Q_{\sigma}^kq_1^{-j+i}q_2^{-\nu_j+\nu_{i-1}+p+1})(1-y^{-1} Q_{\sigma}^k q_1^{-j+i}q_2^{-\mu_j+\nu_{i-1}-p})}{\prod_{i=2}^{e+1}\prod_{j=1}^{c}\prod_{p=0}(1-Q_{\sigma}^k q_1^{-j+i}q_2^{-\nu_j+\nu_{i-1}+p+1})(1- Q_{\sigma}^kq_1^{-j+i}q_2^{-\mu_j+\nu_{i-1}-p})}\bigg)
\nonumber\\ &
\bigg(\prod_{i=2}^{e+1}\prod_{j=1}^{c}\prod_{s=1}^{\mu_j-\nu_j}\nonumber\\&\frac{(1-yq_1^{j-1}q_2^{\nu_j+s-\mu_1-1})}{(1- q_1^{j-1}q_2^{\nu_j+s-\mu_1-1})}\prod_{k=1}\frac{(1-y Q_{\sigma}^k q_1^{j-1}q_2^{\nu_j+s-\mu_{1}-1})(1-y^{-1} Q_{\sigma}^k q_1^{-j+1}q_2^{-\nu_j-s+\mu_{1}+1})}{(1-Q_{\sigma}^kq_1^{j-1}q_2^{\nu_j+s-\mu_{1}-1})(1- Q_{\sigma}^k q_1^{-j+1}q_2^{-\nu_j-s+\mu_{1}+1})}\bigg) \Bigg]\nonumber\\
\eea
For  the values of $r\ge 2$ the compactified web diagram is given  in the figure (\ref{fullcompAnfibP1}), \cite{Iqbal:2015dra}. 
This figure is the web diagram of a CY3-fold described as a resolved $A_{r-1}$ fibration over $\mathbb{P}^1$. We can also see the diagram as obtained by gluing $r-1$ Hirzebruch surfaces. The local geometry of the intersection between $i-1$-th and $i$-th Hirzebruch surfaces is given by the bundle
 $\mathcal{O}(-r+2i+2)\oplus\mathcal{O}(r-2i)\to\mathbb{P}^1$ for $i=1,2,...,r$.
 \\
 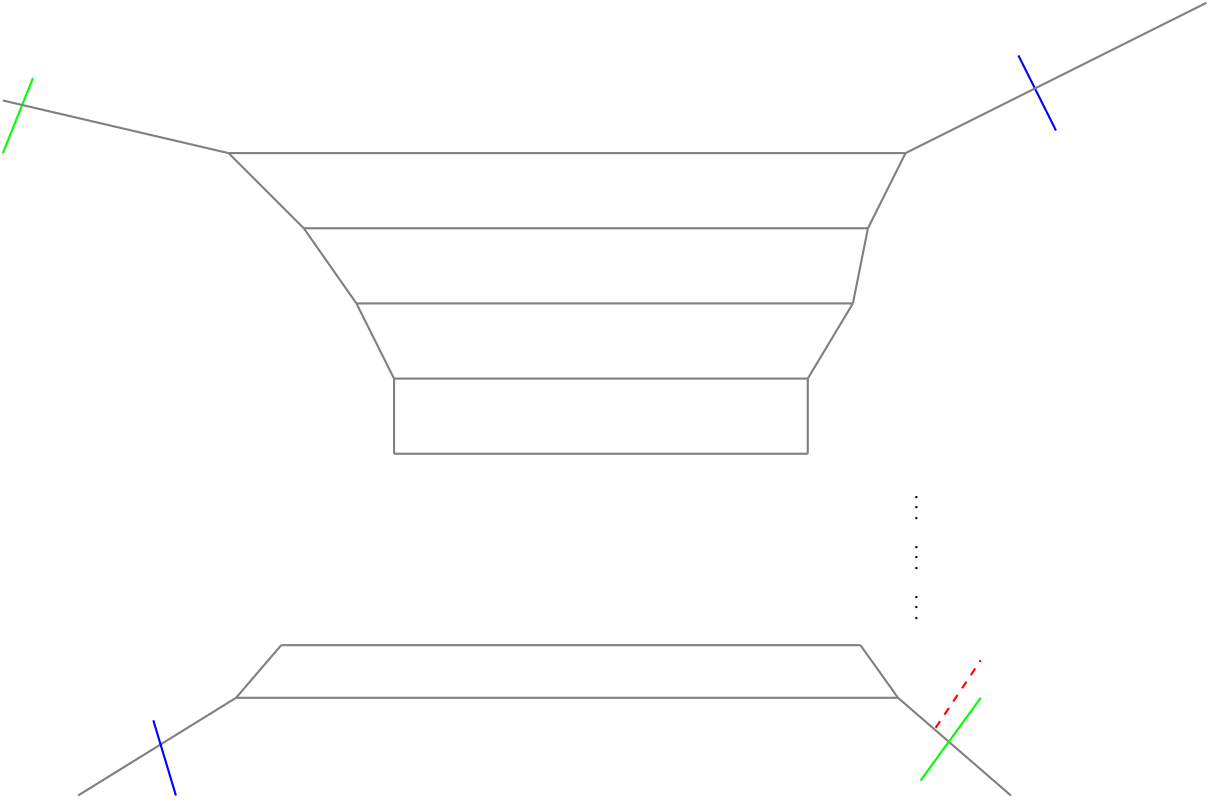
\begin{figure}
\begin{tikzpicture}
\draw[green, thick] (-2,5)-- (-1.6,6);
\draw[blue, thick] (12,5.3) --(11.5,6.3);
\draw[gray, thick] (1,5) -- (10,5);
\draw[gray, thick] (2,4) -- (9.5,4);
\draw[gray, thick] (2,4) -- (1,5);
\draw[gray, thick] (1,5) -- (-2,5.7);
\draw[gray, thick] (9.5,4) --(10,5);
\draw[gray, thick] (10,5) --(14,7);
\draw[gray, thick] (9.5,4) --(9.3,3);
\draw[gray, thick] (2,4) -- (2.7,3);
\draw[gray, thick] (2.7,3) -- (9.3,3);
\draw[gray, thick] (2.7,3) -- (3.2,2);
\draw[gray, thick] (9.3,3) -- (8.7,2);
\draw[gray, thick] (3.2,2) -- (8.7,2);
\draw[gray, thick] (3.2,2) -- (3.2,1);
\draw[gray, thick] (8.7,2) -- (8.7,1);
\draw[gray, thick] (3.2,1) -- (8.7,1);
\vdots
\end{tikzpicture}
\bea
\hspace{7cm}\vdots\nonumber\\
\hspace{7cm}\vdots\nonumber\\
\hspace{7cm}\vdots\nonumber
\eea
\begin{tikzpicture}
\hspace{1cm}\draw[gray, thick] (0.7,1) -- (8.4,1);
\hspace{0cm}\draw[gray, thick] (0.7,1) -- (0.1,0.3);
\hspace{0cm}\draw[gray, thick] (8.4,1) -- (8.9,0.3);
\hspace{0cm}\draw[gray, thick] (0.1,0.3) -- (8.9,0.3);
\hspace{0cm}\draw[gray, thick] (0.1,0.3) -- (-2,-1);
\hspace{0cm}\draw[gray, thick]  (8.9,0.3) -- (10.4,-1);
\hspace{0cm}\draw[blue, thick] (-1,0) -- (-0.7,-1);
\hspace{0cm}\draw[green, thick]  (10,0.3) -- (9.2,-0.8);
\draw[red, thick,dashed] (9.4,-0.1)-- (10,0.8);
\end{tikzpicture}
\caption{fully compactified  web diagram of the the total space of $A_{r-1}$ fibration over $\mathbb{P}^1$}
    \label{fullcompAnfibP1}
\end{figure}
\newpage
 The corresponding elliptic genus for the quiver moduli  $\mathcal{N} (r, k + d, d)$  for $r\ge 2$ is given by
\bea\label{eq:GEG2}
&Z^{6d,quiver}_{r\ge 2,d}=\nonumber\\
&\sum_k Q_{\rho}^k \chi_{ell}(\mathcal{N} (r, k + d, d),q_1,q_2,y,\rho_a,Q_{\sigma})=
\sum_{|\underline{\nu}|=k}Q_{\rho}^k\times\nonumber\\&
\prod_{a,b=1}^r\prod_{(i,j)\in\nu^a}\frac{(1-y\rho_a\rho_b^{-1}q_1^{(\nu^{b})^t_j}q_2^{j-\nu_i^a-1})}{(1-\rho_a\rho_b^{-1}q_1^{(\nu^{b})^t_j}q_2^{j-\nu_i^a-1})}\prod_{k=1}\frac{(1-y Q_{\sigma}^k\rho_a\rho_b^{-1}q_1^{(\nu^{b})^t_j}q_2^{j-\nu_i^a-1})(1-y^{-1} Q_{\sigma}^k\rho_a^{-1}\rho_bq_1^{-(\nu^{b})^t_j}q_2^{-j+\nu_i^a+1})}{(1-Q_{\sigma}^k\rho_a\rho_b^{-1}q_1^{(\nu^{b})^t_j}q_2^{j-\nu_i^a-1})(1-Q_{\sigma}^k\rho_a^{-1}\rho_bq_1^{-(\nu^{b})^t_j}q_2^{-j+\nu_i^a+1})}\nonumber\\&
\prod_{a,b=1}^r\prod_{(i,j)\in\nu^a}\frac{(1-y\rho_a\rho_b^{-1}q_1^{(\nu^{a})^t_j}q_2^{j-\nu_i^b-j})}{(1-\rho_a\rho_b^{-1}q_1^{(\nu^{a})^t_j}q_2^{j-\nu_i^b-j})}\prod_{k=1}\frac{(1-y Q_{\sigma}^k\rho_a\rho_b^{-1}q_1^{(\nu^{a})^t_j}q_2^{j-\nu_i^b-j})(1-y^{-1} Q_{\sigma}^k\rho_a^{-1}\rho_bq_1^{-(\nu^{a})^t_j}q_2^{-j+\nu_i^b+j})}{(1-Q_{\sigma}^k\rho_a\rho_b^{-1}q_1^{(\nu^{a})^t_j}q_2^{j-\nu_i^b-j})(1-Q_{\sigma}^k\rho_a^{-1}\rho_bq_1^{-(\nu^{a})^t_j}q_2^{-j+\nu_i^b+j})}
\nonumber\\&\sum_{\substack{(\mu,\nu)\\|\mu|=|\nu|+d}}
\Bigg[\bigg(\prod_{a=1}^r\prod_{i=1}^{c^a}\prod_{s=1}^{\mu^a_i-\nu^a_i}\rho_a q_a^{1-i}q_2^{-\nu^a_i-s+1}\bigg)\nonumber\\&\times
\nonumber\\&\prod_{a,b=1}^r\bigg(\prod_{i=2}^{e^a+1}\prod_{j=1}^{c^b}\prod_{s=1}^{\mu^b_j-\nu_j^b}\nonumber\\&\frac{(1-y \rho_a\rho_b^{-1}q_1^{j-i}q_2^{\nu_j^b+s-\mu_i^a-1})}{(1- \rho_a\rho_b^{-1}q_1^{j-i}q_2^{\nu_j^b+s-\mu_i^a-1})}\prod_{k=1}\frac{(1-y Q_{\sigma}^k \rho_a\rho_b^{-1}q_1^{j-i}q_2^{\nu_j^b+s-\mu_i^a-1})(1-y^{-1} Q_{\sigma}^k \rho_a^{-1}\rho_bq_1^{-j+i}q_2^{-\nu_j^b-s+\mu_i^a+1})}{(1-Q_{\sigma}^k \rho_a\rho_b^{-1}q_1^{j-i}q_2^{\nu_j^b+s-\mu_i^a-1})(1-Q_{\sigma}^k \rho_a^{-1}\rho_bq_1^{-j+i}q_2^{-\nu_j^b-s+\mu_i^a+1})}\bigg)\nonumber\\ &\prod_{a,b=1}^r
\bigg(\frac{\prod_{i=2}^{e^a+1}\prod_{j=1}^{c^b}\prod_{p=0}(1-y \rho_a\rho_b^{-1}q_1^{j-i}q_2^{\nu_j^b-\nu_{i-1}^a-p-1})(1-y \rho_a\rho_b^{-1}q_1^{j-i}q_2^{\mu_j^b-\nu_{i-1}^a+p})}{\prod_{i=2}^{e^a+1}\prod_{j=1}^{c^b}\prod_{p=0}(1- \rho_a\rho_b^{-1}q_1^{j-i}q_2^{\nu_j^b-\nu_{i-1}^a-p-1})(1- \rho_a\rho_b^{-1}q_1^{j-i}q_2^{\mu_j^b-\nu_{i-1}^a+p})}\nonumber\\&
\frac{\prod_{i=2}^{e^a+1}\prod_{j=1}^{c^b}\prod_{p=0}(1-yQ_{\sigma}^k \rho_a\rho_b^{-1}q_1^{j-i}q_2^{\nu_j^b-\nu_{i-1}^a-p-1})(1-y Q_{\sigma}^k \rho_a\rho_b^{-1}q_1^{j-i}q_2^{\mu_j^b-\nu_{i-1}^a+p})}{\prod_{i=2}^{e^a+1}\prod_{j=1}^{c^b}\prod_{p=0}(1-Q_{\sigma}^k \rho_a\rho_b^{-1}q_1^{j-i}q_2^{\nu_j^b-\nu_{i-1}^a-p-1})(1- Q_{\sigma}^k\rho_a\rho_b^{-1}q_1^{j-i}q_2^{\mu_j^b-\nu_{i-1}^a+p})}\nonumber\\&
\frac{\prod_{i=2}^{e^a+1}\prod_{j=1}^{c^b}\prod_{p=0}(1-y^{-1}Q_{\sigma}^k \rho_b\rho_a^{-1}q_1^{-j+i}q_2^{-\nu_j^b+\nu_{i-1}^a+p+1})(1-y^{-1} Q_{\sigma}^k \rho_b\rho_a^{-1}q_1^{-j+i}q_2^{-\mu_j^b+\nu_{i-1}^a-p})}{\prod_{i=2}^{e^a+1}\prod_{j=1}^{c^b}\prod_{p=0}(1-Q_{\sigma}^k \rho_b\rho_a^{-1}q_1^{-j+i}q_2^{-\nu_j^b+\nu_{i-1}^a+p+1})(1- Q_{\sigma}^k\rho_b\rho_a^{-1}q_1^{-j+i}q_2^{-\mu_j^b+\nu_{i-1}^a-p})}\bigg)
\nonumber\\ &
\bigg(\prod_{a,b=1}^r\prod_{i=2}^{e^a+1}\prod_{j=1}^{c^b}\prod_{s=1}^{\mu^b_j-\nu_j^b}\nonumber\\&\frac{(1-y \rho_a\rho_b^{-1}q_1^{j-1}q_2^{\nu_j^b+s-\mu_1^a-1})}{(1- \rho_a\rho_b^{-1}q_1^{j-1}q_2^{\nu_j^b+s-\mu_1^a-1})}\prod_{k=1}\frac{(1-y Q_{\sigma}^k \rho_a\rho_b^{-1}q_1^{j-1}q_2^{\nu_j^b+s-\mu_{1}^a-1})(1-y^{-1} Q_{\sigma}^k \rho_a^{-1}\rho_bq_1^{-j+1}q_2^{-\nu_j^b-s+\mu_{1}^a+1})}{(1-Q_{\sigma}^k \rho_a\rho_b^{-1}q_1^{j-1}q_2^{\nu_j^b+s-\mu_{1}^a-1})(1- Q_{\sigma}^k \rho_a^{-1}\rho_bq_1^{-j+1}q_2^{-\nu_j^b-s+\mu_{1}^a+1})}\bigg) \Bigg]\nonumber\\
\eea
\section{Conclusions}
We gave expressions for the $\chi_y$ genus and elliptic genus for a quiver moduli space $\mathcal{N} (r, n + d, d)$  described by stable representations of an enhanced ADHM quiver of type (n+d,d,r).
Then  a conjecture is formulated that equates generating function of  $\chi_y$  genus for $r=1$ and $r=2$ with the open topological string partition functions on CY 3-folds given by partially compactified resolved conifold and partially compactified    total space of the bundle $\mathcal{O}(-2,-2)$  of $\mathbb{P}^1\times\mathbb{P}^1$.  Finally it is suggested that the generating function of the elliptic genus may be related to the open topological string partition function on these CY 3-folds whose corresponding web diagrams are fully compactified. To discuss the conjectures for $SU(3)$ and higher rank 5d mass deformed gauge theories, it turns out to be necessary to deal with what are called shifted web diagrams \cite{Bastian:2018fba}.  This is a work in progress. 
\section*{Acknowledgement}
\addcontentsline{toc}{section}{Acknowledgement}
The author would like to thank Amer Iqbal for suggesting the problem and giving various comments. Moreover the support provided by Abdus Salam School of Mathematical Sciences is gratefully acknowledged. 
\begin{appendices}
\section{ $Z_{open}^{ref}(Q_b,Q_f,Q_m,q,t;x)$  on the partially compactified geometry of section (\ref{sectionr2}): Alternate expression }\label{appendix:A}
In section (\ref{sectionr2}) we computed the topological string partition function on the flop of this geometry and then analytically continuted the partition function to the pre-flopped geometry.  
In this appendix  we compute the partition function  using the refined topological vertex without flopping the geometry.
For this case the partially compactified toric diagram of the total space of the bundle $\mathcal{O}(-2,-2)$  of $\mathbb{P}^1\times\mathbb{P}^1$ is given below\\
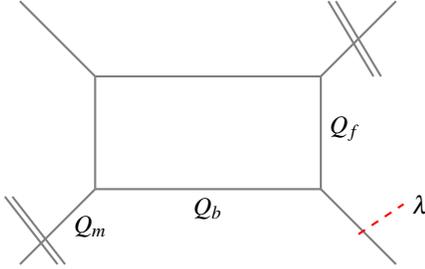
\begin{tikzpicture}
\draw[gray, thick] (2,4) -- (5,4);
\draw[gray, thick] (5,4) -- (6,5);
\draw[gray, thick] (5.8,4) -- (5.2,5);
\draw[gray, thick] (5.7,4) -- (5.1,5);
\draw[gray, thick] (5,4) -- (5,2.5);
\draw[gray, thick] (5,2.5) -- (6,1.5);
\draw[red, thick,dashed] (5.5,1.9)-- (6.1,2.3);
\node [right] at (6.1,2.3) {$\lambda$};
\draw[gray, thick] (5,2.5) -- (2,2.5);
\draw[gray, thick] (2,2.5) -- (1,1.5);
\draw[gray, thick] (1.5,1.5) -- (0.8,2.4);
\draw[gray, thick] (1.6,1.5) -- (0.9,2.4);
\node [right] at (1.6,2) {$Q_m$};
\draw[gray, thick](2,2.5) -- (2,4);
\draw[gray, thick](2,4) -- (1,5);
\node [right] at (5,3.3) {$Q_f$};
\node [below] at (3.5,2.5) {$Q_b$};
\end{tikzpicture}

The building block for $r=2$ corresponding locally to $\mathcal{O}(-2)\oplus \mathcal{O}(0)\to \mathbb{P}^1$   is given by
\bea
Z_{\nu_1\nu_2,\lambda,\rho^t}(q,t,Q_f)=\sum (-Q_f)^{|\mu|}C_{\lambda\mu\nu^t_2}(q,t)C_{\mu^t\rho^t\nu^t_2}(q,t)f_{\mu}(t,q)
\eea
with framing factor $f_{\mu}(t,q)$ and refined topological vertex  $C_{\lambda\mu\nu}(t,q)$ given by
\bea
f_{\mu}(t,q)&=&(-1)^{|\mu|}t^{||\eta^t||/2-|\eta|/2}q^{-||\eta||^2/2+|\eta|/2}\nonumber\\
C_{\lambda\mu\nu}(t,q)&=&(\frac{t}{q})^{\frac{||\mu||^2}{2}}q^{\frac{\kappa(\mu)+||\nu||^2}{2}}\tilde{Z}_{\nu}(t,q)\nonumber\\&\times&\sum_{\eta}(\frac{q}{t})^{\frac{|\eta|+|\lambda|-|\mu|}{2}}s_{\lambda^t/\eta}(t^{-\rho}q^{-\nu})s_{\mu/\eta}(t^{-\nu^t}q^{-\rho})
\eea
The gluing of  the local geometries $\mathcal{O}(-2)\oplus \mathcal{O}(0)\to \mathbb{P}^1$, $\mathcal{O}(0)\oplus \mathcal{O}(-2)\to \mathbb{P}^1$ taking into account the normal-to-the-base directions as well as the compactification of the external leg without brane corresponds to the following amplitude
\bea
Z_{open}^{ref}(Q_f,Q_b,Q_m,q,t;x)&=&\sum (-Q_m)^{|\rho|}(-Q_b)^{|\nu_1|+|\nu_2|}\tilde{f}_{\nu_1^t}(q,t)\tilde{f}_{\nu_2}(t,q)Z_{\nu_1^t\nu_2^t,\emptyset,\rho}(t,q,Q_f)\nonumber\\&\times&Z_{\nu_1\nu_2,\lambda,\rho^t}(q,t,Q_f)\nonumber\\
\eea
Using   the  following skew Schur function identities repeatedly
\bea
\sum_\lambda s_{\lambda/\alpha}(x)s_{\lambda/\beta}(y)&=&\prod_{i,j}(1-x_iy_j)^{-1}\sum s_{\beta/\eta}(x)s_{\alpha/\eta}(y)\nonumber\\
\sum_\lambda s_{\lambda^t/\alpha}(x)s_{\lambda/\beta}(y)&=&\prod_{i,j}(1+x_iy_j)\sum s_{\beta^t/\eta^t}(x)s_{\alpha^t/\eta}(y)
\eea
we get the following expression
\bea
Z_{open}^{ref}(Q_f,Q_b,Q_m,q,t;x)&=&\sum(-Q_b)^{|\nu_1|+|\nu_2|}t^{\frac{||\nu_1^t||^2+||\nu_2^t||^2}{2}}q^{\frac{||\nu_1||^2+||\nu_2||^2}{2}}\nonumber\\&\times&\tilde{f}_{\nu_1^t}(q,t)\tilde{f}_{\nu_2}(t,q)\tilde{Z}_{\nu_1^t}(q,t)\tilde{Z}_{\nu_2^t}(q,t)\tilde{Z}_{\nu_1}(t,q)\tilde{Z}_{\nu_2}(t,q)\nonumber\\&\times&\prod_{i,j}(1-Q_mq^{-\nu_{2i}+j-\frac{1}{2}}t^{i-\frac{1}{2}-\nu_{2j}^t})\prod_{i,j}(1+q^{i-1}t^{-\nu_{1i}^t+\frac{1}{2}}x_j)\nonumber\\&\times&\prod_{i,j}(1-Q_f q^{i-1-\nu_{2j}}t^{j-\nu_{1i}^t})^{-1}\prod_{i,j}(1-Q_f q^{j-\nu_{1i}}t^{i-1-\nu_{2j}^t})^{-1}\nonumber\\&\times&\prod_{i,j}(1-Q_mQ_f q^{j-\frac{1}{2}-\nu_{2i}}t^{i-\frac{1}{2}-\nu_{1j}^t})\prod_{i,j}(1+Q_f q^{i-1}t^{-\nu_{2i}^t+\frac{1}{2}}x_j)\nonumber\\
&\times&\prod_{i,j}(1-Q_mQ_f q^{i-\frac{1}{2}-\nu_{1j}}t^{j-\frac{1}{2}-\nu_{2i}^t})\prod_{i,j}(1-Q_mQ_f^2 q^{j-\frac{1}{2}-\nu_{1j}}t^{i-\frac{1}{2}-\nu_{1j}^t})\nonumber\\&\times&\prod_{i,j}(1+Q_fQ_m\frac{t}{q}t^{-\nu_{2i}^t}q^{i-\frac{1}{2}}x_j)^{-1}
\prod_{i,j}(1+Q_f^2Q_m\frac{t}{q}q^{i-\frac{1}{2}}t^{-\nu_{1i}^t}x_j)^{-1}\nonumber\\
\eea
The gauge theory instanton part of the  partition function that is relevant for the comparison with quiver partition function is given by 
\bea
\tilde{Z}_{open}^{ref}(Q_f,Q_b,Q_m,q,t;x)&=&\sum(-Q_b)^{|\nu_1|+|\nu_2|}t^{\frac{||\nu_1^t||^2+||\nu_2^t||^2}{2}}q^{\frac{||\nu_1||^2+||\nu_2||^2}{2}}\nonumber\\&\times&\tilde{f}_{\nu_1^t}(q,t)\tilde{f}_{\nu_2}(t,q)\tilde{Z}_{\nu_1^t}(q,t)\tilde{Z}_{\nu_2^t}(q,t)\tilde{Z}_{\nu_1}(t,q)\tilde{Z}_{\nu_2}(t,q)\nonumber\\&\times&\frac{\prod_{i,j}(1-Q_mq^{-\nu_{2i}+j-\frac{1}{2}}t^{i-\frac{1}{2}-\nu_{2j}^t})}{\prod_{i,j}(1-Q_mq^{+j-\frac{1}{2}}t^{i-\frac{1}{2}})}\frac{\prod_{i,j}(1-Q_f q^{i-1-\nu_{2j}}t^{j-\nu_{1i}^t})^{-1}}{\prod_{i,j}(1-Q_f q^{i-1}t^{j})^{-1}}\nonumber\\&\times&\frac{\prod_{i,j}(1-Q_f q^{j-\nu_{1i}}t^{i-1-\nu_{2j}^t})^{-1}}{\prod_{i,j}(1-Q_f q^{j}t^{i-1})^{-1}}\frac{\prod_{i,j}(1-Q_mQ_f q^{j-\frac{1}{2}-\nu_{2i}}t^{i-\frac{1}{2}-\nu_{1j}^t})}{\prod_{i,j}(1-Q_mQ_f q^{j-\frac{1}{2}}t^{i-\frac{1}{2}})}\nonumber\\
&\times&\frac{\prod_{i,j}(1-Q_mQ_f q^{i-\frac{1}{2}-\nu_{1j}}t^{j-\frac{1}{2}-\nu_{2i}^t})}{\prod_{i,j}(1-Q_mQ_f q^{i-\frac{1}{2}}t^{j-\frac{1}{2}})}\frac{\prod_{i,j}(1-Q_mQ_f^2 q^{j-\frac{1}{2}-\nu_{1j}}t^{i-\frac{1}{2}-\nu_{1j}^t})}{\prod_{i,j}(1-Q_mQ_f^2 q^{j-\frac{1}{2}}t^{i-\frac{1}{2}})}\nonumber\\&\times&\prod_{i,j}(1+Q_fQ_m\frac{t}{q}t^{-\nu_{2i}^t}q^{i-\frac{1}{2}}x_j)^{-1}
\prod_{i,j}(1+Q_f^2Q_m\frac{t}{q}q^{i-\frac{1}{2}}t^{-\nu_{1i}^t}x_j)^{-1}\nonumber\\&\times&\prod_{i,j}(1+q^{i-1}t^{-\nu_{1i}^t+\frac{1}{2}}x_j)\prod_{i,j}(1+Q_f q^{i-1}t^{-\nu_{2i}^t+\frac{1}{2}}x_j)\nonumber\\
\eea

\section{ $Z_{open}^{ref}(Q_b,Q_f,Q_m,q,t;x)$:  on the geometry  (\ref{flopped}): preferred direction  vertical }\label{appendix:B}
\bea
Z_{open,X_2}^{ref,\nu}(Q_1,Q_2,\tilde{Q}_1,\tilde{Q}_2,Q_\rho,q,t)&=&\sum_{\mu_1,\mu_2,\tilde{\mu_1},\tilde{\mu_2},\rho} (-Q_1)^{|\mu_1|}(-Q_2)^{|\mu_2|}(-\tilde{Q}_1)^{|\tilde{\mu}_1|}(-\tilde{Q}_2)^{|\tilde{\mu}_2|}(-Q_{\rho})^{|\rho|}\nonumber\\&\times&
C_{\tilde{\mu}_2\nu^t\mu_2}(t,q)C_{\tilde{\mu}_2^t\rho\mu_2^t}(q,t)C_{\tilde{\mu}_1\rho^t\mu_1}(t,q)C_{\tilde{\mu}_1^t\emptyset\mu_1^t}(q,t)
\eea

Using the refined topological vertex definition (\ref{eq:RTV2}) and the identities in (\ref{schuridentity})
we get the following expression for the refined open partition function
\bea
&Z&_{open,X_2}^{ref,\nu}(Q_1,Q_2,\tilde{Q}_1,\tilde{Q}_2,Q_\rho,q,t)=\sum_{\mu_1,\mu_2}(-Q_1)^{|\mu_1|}(-Q_2)^{|\mu_2|}\sqrt{t}^{||\mu_1^t||^2+||\mu_2^t||^2+||\nu^t||^2}\sqrt{q}^{||\mu_1||^2+||\mu_2||^2-||\nu||^2}\nonumber\\&\times&\sqrt{\frac{t}{q}}^{|\nu^t|}\tilde{Z}_{\mu_1}(t,q)\tilde{Z}_{\mu_1^t}(q,t)\tilde{Z}_{\mu_2}(t,q)\tilde{Z}_{\mu_2^t}(q,t)
\prod_{i,j}(1-\tilde{Q}_1q^{-\rho_i-\mu_{1j}}t^{-\rho_j-\mu_{1i}^t})\nonumber\\&\times&\prod_{i,j}(1-\tilde{Q}_2q^{-\rho_i-\mu_{2j}}t^{-\rho_j-\mu_{2i}^t})
\prod_{i,j}(1-\tilde{Q}_{\rho}q^{-\rho_i-\mu_{2j}}t^{-\rho_j-\mu_{1i}^t})
\prod_{i,j}(1+\tilde{Q}_1Q_{\rho}q^{-\rho_i-\mu_{2j}}t^{-\rho_j-\mu_{1i}^t})^{-1}\nonumber\\&\times&
\prod_{i,j}(1+\tilde{Q}_2Q_{\rho}q^{-\rho_i-\mu_{1j}}t^{-\rho_j-\mu_{1i}^t})^{-1}\prod_{i,j}(1-\tilde{Q}_1\tilde{Q}_2Q_{\rho}\sqrt{\frac{q}{t}}t^{-\rho_i-\mu_{1j}^t}q^{-\mu_{2i}-\rho_j})\nonumber\\&\times&
(-\tilde{Q}_2\sqrt{t}{q})^{|\nu|}s_{\nu}(-\tilde{Q}_2^{-1}\sqrt{\frac{q}{t}}t^{\mu_2}q^{\rho},-Q_{\rho}^{-1}t^{\mu_1}q^{\rho},\sqrt{\frac{q}{t}}\tilde{Q}_1Q_{\rho}q^{-\rho}t^{-\mu_1^t},q^{-\rho}t^{-\mu_2^t})\nonumber\\
\eea
\section{ $Z_{open}^{ref}(Q_b,Q_f,Q_m,q,t;x)$:  on the geometry  (\ref{flopped}): preferred direction  diagonal }\label{appendix:C}
\bea
Z_{open,X_2}^{ref,\nu}(Q_1,Q_2,\tilde{Q}_1,\tilde{Q}_2,Q_\rho,q,t)&=&\sum_{\mu_1,\mu_2,\tilde{\mu_1},\tilde{\mu_2},\rho} (-Q_1)^{|\mu_1|}(-Q_2)^{|\mu_2|}(-\tilde{Q}_1)^{|\tilde{\mu}_1|}(-\tilde{Q}_2)^{|\tilde{\mu}_2|}(-Q_{\rho})^{|\rho|}\nonumber\\&\times&
C_{\tilde{\mu}_2\tilde{\mu}_2\nu^t}(t,q)C_{\mu_2^t\tilde{\mu_2}^t\rho}(q,t)C_{\mu_1\tilde{\mu}_1\rho^t}(t,q)C_{\mu_1^t\tilde{\mu}^t_1\emptyset}(q,t)
\eea
Using the refined topological vertex definition (\ref{eq:RTV2}) and the identities in (\ref{schuridentity})
we get the following expression for the refined open partition function
\bea
&Z&_{open,X_2}^{ref,\nu}(Q_1,Q_2,\tilde{Q}_1,\tilde{Q}_2,Q_\rho,q,t)=\sum_{\rho}(-Q_{\rho})^{|\mu_\rho|}\tilde{Z}_{\mu_\rho^t}(t,q)\tilde{Z}_{\nu^t}(t,q)\tilde{Z}_{\mu_\rho}(q,t)\sqrt{t}^{||\mu_\rho||^2}\sqrt{q}^{||\mu_\rho^t||^2+||\nu^t||^2}\nonumber\\&\times&\bigg(\prod_{k=1}(1-(Q_1\tilde{Q}_1)^k)^{-1}\prod_{i,j}\frac{(1-(Q_1\tilde{Q}_1)^{k-1}Q_1t^{-\rho_i}q^{-\mu_{i\rho}^t-\rho_j})(1-(Q_1\tilde{Q}_1)^{k-1}\tilde{Q}_1t^{-\mu_{i\rho}-\rho_j}q^{-\rho_i})}{(1-(Q_1\tilde{Q}_1)^{k}\sqrt{\frac{q}{t}}t^{-\rho_i}q^{-\rho_j})(1-(Q_1\tilde{Q}_1)^{k}\sqrt{\frac{t}{q}}t^{-\mu_{i\rho}-\rho_j}q^{-\rho_i-\mu_{j\rho}^t})}    \bigg)\nonumber\\&\times&
\bigg(\prod_{k=1}(1-(Q_2\tilde{Q}_2)^k)^{-1}\frac{(1-(Q_2\tilde{Q}_2)^{k-1}Q_2t^{-\rho_i-\mu_{j\rho}}q^{-\nu^t_{i}-\rho_j})
(1-(Q_2\tilde{Q}_2)^{k-1}\tilde{Q}_2t^{-\nu_i-\rho_j}q^{-\rho_i-\mu_{j\rho}^t})}
{(1-(Q_2\tilde{Q}_2)^{k}\sqrt{\frac{q}{t}}t^{-\rho_i-\mu_{j\rho}}q^{-\rho_j-\mu_{i\rho}^t})(1-(Q_2\tilde{Q}_2)^{k}\sqrt{\frac{t}{q}}t^{-\nu_i-\rho_j}q^{-\rho_i-\nu^t_{j}})} \bigg)\nonumber\\
\eea
\section{Fixed point formulae for the virtual Euler characteristics, the virtual $\chi_y$ genus and the virtual elliptic genus  }\label{virtuallocalization}
In this appendix we summarise the  treatment of the virtual localisation as discussed in \cite{Fantechi_2010}.
Consider a scheme $X$ that is equivariantly embedded into a nonsingular scheme $Y$ globally. There is a $\mathbb{C}^*$ action on the later. Let $Y^f$ denote the nonsingular fixed point locus in $Y$ and let $X^f$ be defined by $X^f=X\cap Y^f$. Moreover $Y^f$ can be written as a union of irreducible components as $:=\cup_{i}Y_i$ and correspondingly $X_i=X\cap Y_i$. We will denote by $[X_i]^{vir}$ the virtual fundamental class, by $\mathcal{O}_{X_i}^{vir}$ the virtual structure sheaf and by $N_i^{vir}$ the virtual normal bundle of $X_i$. For the vector bundle $V$ defined on $X$\footnote{Strictly speaking $V$ is an element of  the Grothendieck  group  of  the vector  bundles on X.} the virtual pushforward $p_*^{vir}()$ is related to the pushforward $p_*^{}()$   by
\bea\label{eq:impid}
p_*^{vir}(V)=p_*(ch(V)td(T^{vir}_X)\cap [X]^{vir})
\eea
Consider an equivariant lift of $V$ denoted by $\tilde{V}$. We consider the restriction of $\tilde{V}$ to $X_i$  denoted by $\tilde{V}_i$ and a projection $p_i:X_i\to pt$. Here and below we denote by  $\mbox{ch}$ the equivariant Chern character, by $\mbox{td}$ the equivariant Todd genus, by $\sum_{i\ge 0}(-1)^i\Lambda^iB$ the action of  $\Lambda_{-1}$ on the vector bundle B and by $\mbox{Eu}$ the equivariant Euler class.\\
 The virtual Riemann-Roch theorem gives
\bea
\chi^{vir}(X,V)=\int_{[X]^{vir}}\mbox{ch}(V).\mbox{td}(T_X^{vir})=p_*(\mbox{ch}(\tilde{V}).\mbox{td}(T^{vir}_X)\cap[X]^{vir})|_{\epsilon=0}
\eea
where the parameter $\epsilon$ is related to the equivariant lift $\tilde{V}$. Then by applying the localisation formula we get

\bea
p_*(\mbox{ch}(\tilde{V}).\mbox{td}(T_X)^{vir}\cap[X]^{vir})&=&\sum_ip_{i*}\big(\mbox{ch}(\tilde{V}_i)\mbox{td}(T_X^{vir}|_{X_i})/\mbox{Eu}(N_i^{vir})\cap[X]^{vir}_i  \big)\nonumber\\
&=&\sum_ip_{i*}\big(\mbox{td}(T_{X_i}^{vir})\mbox{ch}(\tilde{V_i}/\mbox{ch}(\Lambda_{-1}(N_i^{vir}))^{\vee})\cap[X_i]^{vir}  \big)
\eea
Using the following identities 
\bea
\mbox{td}(T_X^{vir}|_{X_i})&=&\mbox{td}(T_{X_i}^{vir})\mbox{td}(N_i^{vir})\nonumber\\
\mbox{td}(N_i^{vir})&=&\mbox{Eu}(N_i^{vir})/\mbox{ch}(\Lambda_{-1}(N_i^{vir})^{\vee})
\eea
and the eq.(\ref{eq:impid}) we get
\bea\label{eq:impid2}
p_*(\mbox{ch}(\tilde{V}).\mbox{td}(T_X)^{vir}\cap[X]^{vir})&=&\sum_ip_{i*}^{vir}(\tilde{V}_i/\Lambda_{-1}(N_i^{vir})^{\vee})
\eea
By definition 
\bea
\chi^{vir}(X,\tilde{V},\epsilon):=\sum_ip_{i*}^{vir}(\tilde{V}_i/\Lambda_{-1}(N_i^{vir})^{\vee})
\eea
Moreover  the eq.(\ref{eq:impid2}) implies that $\chi^{vir}(X,V,\epsilon)\in Q[[\epsilon]]$ and $\chi^{vir}(X,V)=\chi^{vir}(X,\tilde{V},0)$.\\
Similar manipulations lead to the fixed point formulas for the $chi$-y genus and $elliptic$ genus.

\bea
\chi_{-y}^{vir}(X,V)&=&\big(\sum_i\chi_{-y}^{vir}(X_i,\tilde{V}_i\otimes\Lambda_{-y}(N_i^{vir})^{\vee}/\Lambda_{-1}(N_i^{vir})^{\vee}) \big)|_{\epsilon=0}\nonumber\\
\eea
If we define $n_i=rank(N_i^{vir})$ then
\bea
Ell^{vir}(X;z,\tau)=\big(\sum_iy^{-n_i/2}Ell^{vir}(X_i,\mathcal{E}(N_i^{vir})\Lambda_{-y}(N_i^{vir})/\Lambda_{-1}(N_i^{vir})^{\vee},z,\tau)    \big)|_{\epsilon=0}
\eea
\end{appendices}

\bibliographystyle{plain}
\bibliography{bibliography}
\end{document}